\newif\ifdouble
\doubletrue

\ifdouble
\documentclass[journal, 10pt, twoside]{IEEEtran}
\else
\documentclass[journal, 12pt, draftcls, onecolumn, twoside]{IEEEtran}
\fi

\newif\ifcensure

\usepackage{cite}

\ifCLASSINFOpdf
\usepackage[pdftex]{graphicx}

\graphicspath{{./diagrams/}{./images/}}
\DeclareGraphicsExtensions{.pdf,.jpeg,.png,.eps}
\usepackage{epstopdf}
\else
\fi
\usepackage{amsmath}
\usepackage{algorithmic}

\usepackage{algorithm}

\makeatletter
\newcommand\fs@norules{\def\@fs@cfont{\bfseries}\let\@fs@capt\floatc@ruled
	\def\@fs@pre{}%
	\def\@fs@post{}%
	\def\@fs@mid{\kern3pt}%
	\let\@fs@iftopcapt\iftrue}
\makeatother
\restylefloat{algorithm}

\usepackage{array}

\usepackage[caption=false,font=footnotesize]{subfig}
\usepackage{url}

\usepackage{balance}

\usepackage{color}

\usepackage{amsfonts}
\usepackage{amssymb}

\usepackage{hyperref}

\usepackage{tikz}
\usetikzlibrary{patterns}
\usetikzlibrary{backgrounds}
\usetikzlibrary{calc}
\usetikzlibrary{shapes}
\usetikzlibrary{fit}
\usetikzlibrary{chains}
\usetikzlibrary{arrows}
\usepackage{multirow}
\usepackage{tkz-graph}
\usetikzlibrary{intersections}
\usepackage{tkz-euclide}
\usetkzobj{all}
\usepackage{multirow}
\usepackage{tkz-graph}

\tikzstyle{block} = [draw, rectangle, minimum height=2em, minimum width=0.5em]
\tikzstyle{sum} = [draw, circle, node distance=1cm, fill=white]
\tikzstyle{input} = [coordinate] \tikzstyle{output} = [coordinate]
\tikzstyle{pinstyle} = [pin edge={to-,thin,black}]

\usepackage[normalem]{ulem}

\newcommand{\dbar}{{\bar{d}}}

\begin{document}

	%
	\title{Iterative Equalization with Decision Feedback based on Expectation Propagation}
	%
	%
	%
	
	\author{Serdar~\c{S}ah\.{i}n,
		Antonio~Maria~Cipriano,
		Charly~Poulliat
		and~Marie-Laure~Boucheret%
		\thanks{Manuscript received November 16, 2017; revised April 05, 2018 and accepted on May 24, 2018. }
		\thanks{S. \c{S}ah\.{i}n is with both Thales Communications \& Security and IRIT/INPT-ENSEEIHT (e-mail: serdar.sahin@thalesgroup.com)}%
		\thanks{A. M. Cipriano is with Thales, 4 Av. des Louvresses, 92230, Gennevilliers, France (e-mail: antonio.cipriano@thalesgroup.com)}%
		\thanks{C. Poulliat and M.-L. Boucheret are with IRIT / INP Toulouse -ENSEEIHT, 2 Rue Charles Camichel, 31000, Toulouse, France (e-mails: $\lbrace\text{charly.pouillat, marie-laure.boucheret}\rbrace$@enseeiht.fr)}%
	}
	
	%
	%

	\markboth{To appear on IEEE Journal on Transactions on Communications - May 2018}%
	{\c{S}ah\.{i}n \MakeLowercase{\textit{et al.}} -  Iterative Equalization with Decision Feedback based on Expectation Propagation}
	%

	\IEEEpubid{
		\begin{minipage}{\textwidth}\ \\[12pt] \centering
			\copyright IEEE. Personal use of this material is permitted. However, permission to reprint/republish this material for advertising or promotional purposes or for creating new collective works for resale or redistribution to servers or lists, or to reuse any copyrighted component of this work in other works must be obtained from the IEEE. Final version: https://ieeexplore.ieee.org/document/8371591/
		\end{minipage}
		 }


	\maketitle
	
	\begin{abstract}
		This paper investigates the design and analysis of minimum mean square error (MMSE) turbo decision feedback equalization (DFE), with expectation propagation (EP), for single carrier modulations. Classical non iterative DFE structures have substantial advantages at high data rates, even compared to \textit{turbo} linear equalizers - interference cancellers (LE-IC), hence making turbo DFE-IC schemes an attractive solution. In this paper, we derive an iterative DFE-IC, capitalizing on the use of soft feedback based on expectation propagation, along with the use of prior information for improved filtering and interference cancellation. This DFE-IC significantly outperforms exact turbo LE-IC, especially at high spectral efficiency, and also exhibits various advantages and performance improvements over existing variants of DFE-IC. 
		The proposed scheme can also be self-iterated, as done in the recent trend on EP-based equalizers, and it is shown to be an attractive alternative to linear self-iterated receivers.
		For time-varying (TV) filter equalizers, an efficient matrix inversion scheme is also proposed, considerably reducing the computational complexity relative to existing methods. 
		Using finite-length and asymptotic analysis on a severely selective channel, 
		the proposed DFE-IC is shown to achieve higher rates than known alternatives, with better waterfall thresholds and faster convergence, while keeping a similar computational complexity.	
	\end{abstract}
	

	%
	\IEEEpeerreviewmaketitle

	\section{Introduction}

	\ifdouble
	\IEEEPARstart{C}OMMUNICATION systems operating on wide-band channels suffer from inter-symbol interference (ISI), which can be mitigated with an appropriate transceiver design. In particular, for wireless systems where the throughput requirements increase at each new generation, more effective receivers are needed in order to maintain robust data links.
	\else
	Communication systems operating on wide-band channels suffer from inter-symbol interference (ISI), which can be mitigated with an appropriate transceiver design. In particular, for wireless systems where the throughput requirements increase at each new generation, more effective receivers are needed in order to maintain robust data links.
	\fi

	With the discovery of turbo-codes, iterative processing principles were extended to joint detection and decoding techniques via soft-input soft-output (SISO) receivers which use prior information provided by the channel decoder, to further reduce detection errors. Although early turbo equalization techniques, such as maximum a posteriori (MAP) detector using BCJR estimation \cite{douillard_iterative_1995, bahl_optimal_1974, bauch_1994_comparisonSISOturbodetection}, can operate near the channel capacity with properly designed coding schemes, their operational complexity significantly increases for large channel delay spread or with high modulation orders. 
	Consequently, finite impulse response (FIR) filter-based turbo equalizers with lowered computational complexity have been proposed. 
	These structures can be categorized into three groups with regards to its filter updates depending on prior information. Other kinds of adaptive FIR receivers are out of this paper's scope.  Time-invariant (TI) structures  update their filters only once at each packet reception, using the available channel state. Iteration-variant (IV) equalizers are updated at each turbo iteration by additionally using the overall prior information. Time-varying (TV) structures update their filters at each symbol, using both symbol-wise prior information and channel states, making them particularly suitable for doubly selective channels, where the impulse response varies over time. 
	
	\IEEEpubidadjcol
	The first FIR turbo structure, proposed by Laot et al. \cite{glavieux_turbo_1997}, uses a time-invariant interference canceller \cite{gersho_1981_interferenceCanceller}, and an application to IV filtering appeared in \cite{fijalkow_improved_2000, roumy_egalisation_2000}. Further extension to TV equalization is provided in \cite{wang_1999_iterativeSoftICCDMA} and a formal framework presented in \cite{boutros_2002_iterativeMUDCDMA} derive these receivers from the MAP criterion.

	An alternative approach formalized by T\"{u}chler et. al \cite{tuchler_minimum_2002} consists in designing a TV adaptive LE, by using statistics conditioned on prior information, while solving the MMSE criterion. This structure has been applied to high-order modulations, time-varying channels and to IV, TI, frequency domain structures for lower complexity, and also to multi-user detection for multiple input-multiple output systems \cite{dejonghe_turbo-equalization_2002, tuchler_linear_2001,otnes_iterative_2004, visoz_2005_new}.
	
	Equivalence of these approaches was shown in \cite{le_bidan_turbo-equalization_2003}, making the TV MMSE LE-IC the most widespread reference. Although turbo LE-IC brings significant improvements over classical filtering, it falls far behind classical DFE \cite{belfiore_79_DFE, cioffi_2008_equalization} at high spectral efficiency operating points.
	Oppositely, at lower rates, turbo LE-IC is near capacity-achieving while DFE performs poorly\footnote{These facts are also shown in subsection \ref{subsec_asymanalysis}, in Fig. \ref{fig_effSpecProakisCIntro}.}.
	
	This paper addresses the design of iterative time-domain TV DFE-IC equalizers, i.e. FIR receivers where prior information and a symbol-wise decision feedback is  respectively used on anti-causal and causal symbols, to improve equalization. These receivers are of interest for applications where doubly-selective channels are involved, such as HF communications \cite{elgenedy_2013_iterativeDFEforHF}.

	\subsection{Related Work}

	There exist several prior works on DFE-IC. Proposals mainly differ with the nature of decision feedback, and with the filter updating method. Besides, recent complex receivers use DFEs as constituent elements for concatenated equalizers. 
	Hence, for clarity, we propose to classify related works in three sub-categories.

	\subsubsection{Iterative Hard DFE-IC}
	Among hard feedback structures, DFE-IC in \cite{tuchler_turbo_2002} 
	is a classical DFE that uses prior information for IC on anti-causal symbols. This structure is known for its error propagation issues which makes its TV form even less efficient than TI LE, and its extrinsic information transfer (EXIT) analysis yields contradictory results \cite[Fig. 14]{tuchler_turbo_2002}.
	In \cite{jeong_soft-soft-out_2011}, the previous structure  
	is enhanced with a powerful soft demapper that uses the distribution of residual ISI sequences for symbol detection. This modified structure outperforms turbo LE-IC, but this residual ISI distribution is very difficult to derive even in the simple BPSK case. A more practical solution, proposed in \cite{jeong_turbo_2010}, consists in approximating the residual ISI at the DFE-IC output to an additive white Gaussian noise (AWGN), which simplifies the demapper. While this solution challenges TV LE-IC on BPSK, its extension to multilevel modulations has not been explored so far. To the authors' knowledge, this is the only DFE-IC outperforming exact TV LE-IC in the reference scenario of Proakis-C channel with BPSK symbols.
	DFE-ICs in \cite{jeong_turbo_2010,jeong_soft-soft-out_2011} were later used as constituent elements for more advanced receivers such as bi-directional DFE, or structures obtained by parallel concatenation of FIRs \cite{jeongMoon_2013_selfiteratingSoftEqualizer}.
	
	\subsubsection{Iterative Soft DFE-IC}
	Literature on turbo soft DFE-IC is more diverse; although feedback is mostly based on the posterior distribution, there is no common strategy for evaluating its variance \cite{lopes_soft-feedback_2006,tao_2016_low,trajkovic_turbo_2005}.
	Such iterative structure is first presented in \cite{balakrishnan_mitigation_1999}, where various TI DFE with soft feedback are evaluated with a perfect decision hypothesis, within a sub-optimal receiver using hard decoding. In particular, it is seen that soft feedback mitigates to some extent error propagation, despite ignoring decision errors in filter computation.
	Another notable structure is the IV 
	soft interference canceller in \cite{lopes_soft-feedback_2006}; using both prior and posterior LLRs for filtering and for interference cancellation with BPSK, this scheme significantly outperforms IV LE-IC, but it requires stochastic methods for estimating the correlation properties of posterior LLRs. 
	Several other IV soft feedback structures exist \cite{trajkovic_turbo_2005,lou_soft-decision_2011}, with alternative heuristics for feedback quality assessment. 
	Structural comparison of IV schemes using posterior feedback is given in \cite{tao_2016_low}, extending \cite{lopes_soft-feedback_2006} and \cite{lou_soft-decision_2011} to higher order modulations, but requiring new heuristics with LE-IC pre-equalization for filter computation. These approaches have drawbacks due to their limitations in usable constellations
	\cite{lopes_soft-feedback_2006,trajkovic_turbo_2005,lou_soft-decision_2011}, or due to the sub-optimality of heuristics used in filter computation \cite{lopes_soft-feedback_2006,lou_soft-decision_2011,tao_2016_low}. 
	Indeed, IV structures need static statistics of its soft feedback for computing its filters, which requires approximations. 
	
	\begin{table*}[t]
		\caption{Classification of Constituent FIR Turbo Equalizers vs. the Usage of Prior Information.}
		\label{table:refclass}
		\centering
		\begin{tabular}{|c||c|c|c||c|c|c|c|}
			\hline
			&  \multicolumn{3}{c||}{Linear Structure} & \multicolumn{4}{c|}{Decision Feedback Structures} \\ \hline
			Update Type & TI & IV & TV & \multicolumn{1}{c||}{Dec. Type} & TI & IV & TV \\ \cline{1-8}
			\multirow{3}{*}{References} & \multirow{3}{*}{\parbox{1cm}{\centering \cite{glavieux_turbo_1997,tuchler_turbo_2002}}} & \multirow{3}{*}{\parbox{1.75cm}{\centering \cite{fijalkow_improved_2000,roumy_egalisation_2000,boutros_2002_iterativeMUDCDMA,tuchler_minimum_2002,tuchler_linear_2001,visoz_2005_new,le_bidan_turbo-equalization_2003}}} & \multirow{3}{*}{\parbox{1.5cm}{\centering \cite{tuchler_minimum_2002,tuchler_turbo_2002,boutros_2002_iterativeMUDCDMA,wang_1999_iterativeSoftICCDMA,dejonghe_turbo-equalization_2002,otnes_iterative_2004}}} & \multicolumn{1}{c||}{Hard} & \cite{jeong_soft-soft-out_2011,jeong_turbo_2010} & \cite{jeong_soft-soft-out_2011,jeong_turbo_2010} & \cite{tuchler_turbo_2002,jeong_soft-soft-out_2011,jeong_turbo_2010}  \\ \cline{5-8}
			& & & & \multicolumn{1}{c||}{Soft APP} & \cite{balakrishnan_mitigation_1999} & \cite{tao_2016_low,lopes_soft-feedback_2006,lou_soft-decision_2011,tao_2011_enhanced} & {\textbf{Proposed}} \\ \cline{5-8}
			& & & & \multicolumn{1}{c||}{Soft EP} & &  & {\textbf{Proposed}} \\ \cline{1-8} 
		\end{tabular}
	\end{table*}
	
	Time-varying soft posterior feedback structures do not have such issues; they can update their filters after each symbol is detected, as it had been done for MIMO receivers in \cite{choi_2008_improved}. In equalization, the structure closest to \cite{choi_2008_improved} is a block-feedback turbo DFE in \cite{tao_2011_enhanced}, which updates its filters every $P$ symbols. A classification of the references above is given in Table \ref{table:refclass}.
	
	\subsubsection{Receivers based on Expectation Propagation}
	There is a recent renewal of interest in iterative equalization, brought by the use of  an approximate statistical inference method, namely expectation propagation (EP) \cite{minka_2001_expectationpropagation}.  
	This technique can be used as a message passing algorithm, which extends the loopy belief propagation (BP) by using exchange of expectations. 
	When EP is used with probability density functions (PDF) belonging to the exponential family, it is possible to compute an extrinsic message passed from the demapper to the equalizer. 
	
	This paradigm has already been used in channel decoding \cite{walsh_2006_iterativedecodingwithEP}, and in receiver design with MIMO receivers \cite{senstAscheid_2011_frameworkEP_MMSEMIMO}, block linear equalizers \cite{santosMurilloFuentes_2017_EPBLE} and Kalman smoothers \cite{sunFleury_2015_iterativeRecvCombinedBPEP, santosMurillosFuentes_2017_smoorthingEP}. 
	In particular, a concomitant work has recently extended these schemes to FIR with a self-iterated LE-IC \cite{santosMurilloFuentes_2017_EPnuBLE}. In \cite{sahinCipriano_2018_FDSILEEP}, EP was applied on multivariate white Gaussian distributions to derive a low-complexity self-iterated frequency domain equalizer.
	The receivers above use EP in a parallel interference cancellation scheduling through self-iterations, i.e. the whole data block is detected, and then detection process is repeated using EP feedback from the demapper. These structures are not decision feedback structures as in \cite{belfiore_79_DFE}, which are natural successive interference cancellers. 
	
	Hence, in this paper, we propose to derive a \emph{DFE-IC EP} exploiting the successive interference cancellation schedule of DFE-IC to operate on an EP-based soft feedback. Moreover, we combine this serial detection framework with an outer loop, as in prior work on EP, to obtain a \emph{self-iterated DFE-IC EP}.
	A low complexity matrix inversion strategy for TV FIR structures is also derived, significantly reducing the computational complexity difference between DFE-IC and LE-IC.

	\subsection{Contributions and Paper Outline}
	
	The main contributions of this paper are as follows:
	\begin{itemize}
		\item[$\bullet$] A novel time-varying DFE-IC algorithm, using EP to update its filters, and to cancel residual ISI, is proposed. It outperforms other constituent FIR receivers known to the authors, while providing an overall efficient complexity-performance trade-off.
		\item[$\bullet$] DFE-IC EP is extended to a self-iterated structure, and compared to prior work on self-iterated EP receivers.
		\item[$\bullet$] Well-known hard \cite{tuchler_turbo_2002,jeong_turbo_2010} or sub-optimal \cite{tao_2016_low,tao_2011_enhanced} DFE-IC proposals  are extended to TV structures with soft posterior feedback, by using MMSE Bayesian estimators.
		\item[$\bullet$] Analytical and asymptotic analysis of DFE-IC is carried out on a highly selective deterministic channel. Performance and computational complexity comparison between LE-IC and different DFE-IC structures is provided.
		\item[$\bullet$] A new recursive matrix inversion strategy for TV equalizers is exposed. Compared to the iterative algorithm in \cite{tuchler_minimum_2002}, it brings between 30\% (for long data blocks) and 75\% (for shorter blocks) complexity reduction for LE-IC.
	\end{itemize}

	The remainder of this paper is organized as follows. The considered BICM communication scheme and the generic FIR receiver model are described in section \ref{sec:modelling}.  Section \ref{sec:ep_modelling} proposes a factor graph model for the system and applies the expectation propagation framework to derive the proposed equalizer in subsection \ref{ssec:dfeic_ep}. 
	A novel matrix inversion strategy is detailed in section \ref{sec:matrixinvcholesk} for reducing TV equalization complexity.
	Section \ref{sec:dfe} extends prior work on DFE-IC to the state-of-the-art and compares with the proposed DFE-IC EP. In section \ref{sec:self}, DFE-IC EP is self-iterated, and compared with several existing self-iterated EP receivers.
	
	\subsection{Notations}
	Bold lowercase letters are used for vectors: let $\mathbf{u}$ be a $N \times 1$ vector, then $u_n, n=0,\dots,N-1$ are its entries. 
	Capital bold letters denote matrices: for a given $N \times M$ matrix $\mathbf{A}$, $[\mathbf{A}]_{n,:}$ and $[\mathbf{A}]_{:,m}$ respectively denote its $n^\text{th}$ row and $m^\text{th}$ column, and $a_{n,m}=[\mathbf{A}]_{n,m}$ is the entry $(n,m)$.
	
	$\mathbf{I}_N$ is the $N\times N$ identity matrix, $\mathbf{0}_{N,M}$ and $\mathbf{1}_{N,M}$ are respectively all zeros and all ones $N\times M$ matrices. $\mathbf{e}_n$ is the $N\times 1$ indicator whose only non-zero entry is $e_n=1$.
	Operator $\textbf{Diag}(\mathbf{u})$ denotes the diagonal matrix whose diagonal is defined by $\mathbf{u}$. 
	$\mathbb{R}, \mathbb{C}$, and $\mathbb{F}_k$ are respectively the real field, the complex field and a Galois field of order $k$.
	Let $x$ and $y$ be two random variables, then $\mu_x=\mathbb{E}[x]$ is the expected value, $\sigma_x^2=\text{Var}[x]$ is the variance and $\sigma_{x,y}=\text{Cov}[x,y]$ is the covariance. The probability of $x$ taking a value $\alpha$ is $\mathbb{P}[x=\alpha]$, and probability density functions (PDF) are denoted as $p(\cdot)$.  
	If $\mathbf{x}$ and $\mathbf{y}$ are random vectors, then we define vectors $\pmb{\mu}_\mathbf{x}=\mathbb{E}[\mathbf{x}]$ and $\pmb{\sigma}_\mathbf{x}^2=\text{Var}[\mathbf{x}]$, the covariance matrix $\mathbf{\Sigma}_{\mathbf{x},\mathbf{y}}=\textbf{Cov}[\mathbf{x,y}]$ and we note $\mathbf{\Sigma}_{\mathbf{x}}=\textbf{Cov}[\mathbf{x,x}]$.
	$\mathcal{CN}(\mu_x,\sigma_x^2)$ denotes the circularly-symmetric complex Gaussian distribution of mean $\mu_x$ and variance $\sigma_x^2$, and  $\mathcal{B}(p)$ denotes the Bernoulli distribution with a success probability of $0\leq p \leq 1$.

	\section{System Model}\label{sec:modelling}
	\subsection{Transmission Over a Multipath Channel}
	
	We consider a single carrier transmission using a bit-interleaved coded modulation (BICM) scheme. Let $\mathbf{b}\in\mathbb{F}_2^{K_b}$ be a binary information packet of length $K_b$ bits. 
	A channel encoder maps $\mathbf{b}$ into a codeword $\mathbf{c}\in\mathbb{F}_2^{K_c}$, with a code rate $R_c=K_b/K_c$, which is then
	interleaved to give a data block $\mathbf{d}\in \mathbb{F}_2^{K_c}$. 
	A memoryless mapping $\varphi$ associates $\mathbf{d}$ to the symbol block of length $K$, denoted $\mathbf{x}\in \mathcal{X}^K$, where the constellation $\mathcal{X} \subset \mathbb{C}$ has $M$ elements. The $q$-word associated to a symbol is denoted $\mathbf{d}_k=[\mathbf{d}]_{qk:q(k+1)-1}$, and $\varphi_j^{-1}(x_{k})$ and $d_{k,j}$ denote the value of the $j^\text{th}$ bit labelling the $k^\text{th}$ symbol $x_{k}$, i.e. $d_{kq+j}$. 
	We assume the constellation has zero mean, and has an average symbol power of $\sigma^2_x$, with equiprobable symbols.

	For the sake of clarity, only the single user, single input-single output $T$-spaced (symbol spaced) equalization problem is considered. 
	The channel is modelled at the base-band as an equivalent $L$-tap linear time-varying filter $\mathbf{h}[k]=[h_{k,L-1}, h_{k,L-2} \dots h_{k,0}]$, $k$ being the time index, and where pulse shaping and transceiver filters are accounted for. 
	
	The signal going through the channel is then affected by thermal noise $w_{k}$ at the receiver side, and assuming a perfect channel state information, ideal time and frequency synchronization and the absence of inter-block interference (IBI), the base-band received samples are given by:
	\begin{equation}
		y_{k} = \textstyle \sum_{l=0}^{L-1}{h_{k,l} x_{k-l}} + w_{k},
	\end{equation}
	where $k=0,1,\dots,K+L-2$, and $x_{k},\, k < 0$ and $k > K$ are set to $0$. These assumptions can be satisfactorily approached in practice with the use of a unique-word signalling scheme, among other options, to jointly enable channel estimation and the IBI removal.
	The noise is modelled as $w_{k}\sim \mathcal{CN}(0, \sigma_w^2)$, i.e. its real and imaginary parts are real independent zero mean Gaussian random processes with $\sigma_w^2/2$ variance each. 
	The transmission can be rewritten as:
	\begin{equation}
		\mathbf{y} = \mathbf{H}\mathbf{x}+\mathbf{w},
	\end{equation}
	with $\mathbf{y}=[y_0, \dots, y_{K+L-2}]^T$, $\mathbf{w}=[w_0, \dots, w_{K+L-2}]^T$, $\mathbf{x}=[x_{-L+1}, \dots, x_{K+L-2}]^T$ and $\mathbf{H}$ is the $(K+L-1)\times (K+2L-2)$ matrix whose $k^\text{th}$ row is $\left[\mathbf{0}_{1,k-1}, \mathbf{h}[k], \mathbf{0}_{1,K+L-1-k}\right]$, $k=1,\dots,K+L-1$.

	\subsection{On MMSE FIR Equalization}\label{sec:ep_mmsefir}
	
	FIR structures can be modelled by windowed processes; applying a sliding window $[-N_p,N_d]$ on the observation vector $\mathbf{y}$, we define $\mathbf{y}_k=[y_{k-N_p},\dots,y_{k+N_d}]^T$.
	$N_p$ and $N_d$ are respectively the number of pre-cursor and post-cursor samples, and we denote $N\triangleq N_p+N_d+1$, and $N_p'\triangleq N_p+L-1$ to simplify notations.  
	Then, using the same window on $\mathbf{w}$, and $[-N_p',N_d]$ on $\mathbf{x}$, the channel model becomes
	\begin{equation}
		\mathbf{y}_k = \mathbf{H}_k\mathbf{x}_k+\mathbf{w}_k, \label{eq_slidingwindow}
	\end{equation}
	with $\mathbf{H}_k = [\mathbf{H}]_{k-N_p \,:\, k+N_d,\: k-N_p' \,:\, k+N_d}$, for $k=0,\dots,K-1$. 
	
	Below, a generic structure of an unbiased MMSE FIR receiver is given for comparing different structures and their dynamics in the remainder of the paper.
	Prior estimates on $\mathbf{x}$ with means $\mathbf{\bar{x}}_k^{\textbf{fir}}\triangleq [\bar{x}_{k-N_p'}^{\text{fir}},\dots,\bar{x}_{k+N_d}^{\text{fir}}]$ and variances $\mathbf{\bar{v}}_k^{\textbf{fir}}\triangleq [ \bar{v}_{k-N_p'}^{\text{fir}},\dots,\bar{v}_{k+N_d}^{\text{fir}}]$ are used for interference cancellation. Then denoting its output estimate on $x_k$ as $x_k^e$, and the variance of the residual interference and noise as $v_k^e$, with
	\begin{eqnarray}
		\begin{array}{ll}
			{x}^e_k &= \mathbf{f}_k^{\textbf{fir}}{}^H \mathbf{y}_k + g_k^{\text{fir}}\\
			v^e_k &= 1/\xi^{\text{fir}}_k  - \bar{v}^{\text{fir}}_k
		\end{array}
		,\, \text{ } 
		\begin{cases}
			\mathbf{f}_k^{\textbf{fir}} \triangleq \mathbf{\Sigma}^{\textbf{fir}}_k{}^{-1}\mathbf{h}_k/\xi^{\text{fir}}_k, \\
			g_k^{\text{fir}} \triangleq \bar{x}^{\text{fir}}_k - \mathbf{f}_k^{\textbf{fir}}{}^{H}\mathbf{H}_k\mathbf{\bar{x}}^{\textbf{fir}}_k,\\
			\xi^\text{fir}_k \triangleq \mathbf{h}_k^H\mathbf{\Sigma}^{\textbf{fir}}_k{}^{-1}\mathbf{h}_k,
		\end{cases}
		\label{eq_fir_model}
	\end{eqnarray}
	where $\mathbf{\Sigma}^{\textbf{fir}}_k \triangleq k_w\sigma_w^2\mathbf{I}_N + \mathbf{H}_k\mathbf{\bar{V}}^\textbf{fir}_k\mathbf{H}_k^H$, $\mathbf{\bar{V}}^\textbf{fir}_k \triangleq\textbf{diag}(\mathbf{\bar{v}}^\textbf{fir}_k)$, $\mathbf{h}_k \triangleq\mathbf{H}_k\mathbf{e}_k$ and $k_w=1/2$, when signals with one real degree of freedom are used (e.g. $\mathcal{X}$ is BPSK), and otherwise $k_w=1$ \cite{cioffi_2008_equalization}.
	A proof of these relationships is in Appendix \ref{sec_app_mmse_fir}. 
	
	Note that $\mathbf{\bar{x}}_k^{\textbf{fir}}$ and $\mathbf{\bar{v}}_k^{\textbf{fir}}$ completely characterize such receivers.
	When $\mathbf{\bar{x}}_{k'}^{\textbf{fir}}$ and $\mathbf{\bar{v}}_{k'}^{\textbf{fir}}$ are independent of $x^e_{k}, v^e_{k}, \forall k',k$, we call this receiver a LE-IC, and when $\bar{x}_{k'}^{\text{fir}}$ and ${\bar{v}}_{k'}^{\text{fir}}$ are dependent on $x^e_{k}, v^e_{k}$, $\forall k'<k$, we refer to it as a DFE-IC.
	
	\section{Receiver Design with Expectation Propagation}\label{sec:ep_modelling}
	
	This section focuses on the design of a FIR receiver that approximates the posterior probability distribution on $x_k$ using an EP-based message passing on the system factor graph.
	
	\subsection{Factor Graph Model for FIR Receivers}\label{ssec:ep_factorgraph}

	The optimal joint MAP receiver satisfies the MAP criterion $\hat{\mathbf{b}}=\max_\mathbf{b} p(\mathbf{b}\vert \mathbf{y})$, where, assuming i.i.d. information bits, the posterior PDF can be factorized as follows
	\begin{equation}
		p(\mathbf{b} \vert \mathbf{y}) = p(\mathbf{b}, \mathbf{d}, \mathbf{x} \vert \mathbf{y}) \propto \underbrace{p(\mathbf{y}\vert\mathbf{x})}_\text{channel} \underbrace{p(\mathbf{x}\vert \mathbf{d})}_\text{mapping} \underbrace{p(\mathbf{d\vert b})}_\text{encoding}.\label{eq_postb}
	\end{equation}
	This density can be further factorized by using:
	\begin{itemize}
	\item[-] the memoryless mapping: $p(\mathbf{x}\vert \mathbf{d}) = \prod_{k=0}^{K-1}p( x_k\vert \mathbf{d}_k)$,
		\item[-] the independence assumption in BICM encoding: $p(\mathbf{d\vert b}) = \prod_{k=0}^{K-1}\prod_{j=0}^{q-1}p(d_{k,j})$, 
	\end{itemize}
	where $p(d_{k,j})~\triangleq~p(d_{k,j}\vert \mathbf{b})$ is a probability mass function (PMF) which is seen as a Bernoulli-distributed prior constraint provided by the decoder, from the receiver's point of view. 
	
	The ``channel" factor in (\ref{eq_postb}) creates constraints between the whole block of received baseband samples and the transmitted symbols, however to derive a reduced complexity FIR receiver which estimates $x_k$ and $\mathbf{d}_k$, the windowed model in (\ref{eq_slidingwindow}) is needed.
	The FIR approximation posterior is
	\ifdouble
	\begin{equation}
		\label{eq_postfir}
		\begin{split}
			\textstyle p\left(\mathbf{\dbar}_k, \mathbf{x}_k \vert \mathbf{y}_k\right) \propto \prod_{k'=k-N_p'}^{k+N_d}  & \textstyle p(\mathbf{y}_k\vert \mathbf{x}_k) p(x_{k'}\vert \mathbf{d}_{k'})\\
			\textstyle & \textstyle \prod_{j=0}^{q-1}p(d_{k',j}), 
		\end{split}
	\end{equation}
	\else
	\begin{equation}
		\textstyle p\left(\mathbf{\dbar}_k, \mathbf{x}_k \vert \mathbf{y}_k\right) \propto p(\mathbf{y}_k\vert \mathbf{x}_k) \prod_{k'=k-N_p'}^{k+N_d} p(x_{k'}\vert \mathbf{d}_{k'}) \prod_{j=0}^{q-1}p(d_{k',j}), \label{eq_postfir}
	\end{equation}
	\fi
	where $\mathbf{\dbar}_k=\mathbf{d}_{k-N_p-L+1:k+N_d}$.
	Note that working with $p\left(\mathbf{\dbar}_k, \mathbf{x}_k \vert \mathbf{y}\right)\approx \textstyle p\left(\mathbf{\dbar}_k, \mathbf{x}_k \vert \mathbf{y}_k\right)$ is not the only option for estimating $x_k$. Indeed $x_k$ can be estimated through inference on $\mathbf{x}_{k'}$, with $k'=k-N_d,\dots,k+N_p'$, but by selecting $\mathbf{x}_k$, this option is indirectly translated to the choice of window parameters, which is a common aspect of FIR equalizers.

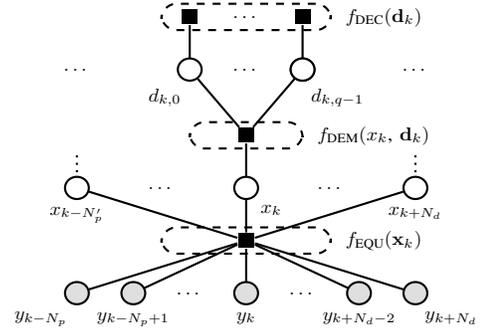
\begin{figure}[!t]
		\centering
		\begin{tikzpicture}[thick, xscale=0.75, yscale=0.7, every node/.style={xscale=0.75, yscale=0.7}]
		\tikzstyle{latent} = [circle,fill=white,draw=black,inner sep=0.15pt,
		minimum size=12pt, node distance=0]
		\tikzstyle{obs} = [latent,fill=gray!25]
		\tikzstyle{const} = [rectangle, inner sep=0pt, node distance=1]
		\tikzstyle{factor} = [rectangle, fill=black,minimum size=8pt, inner
		sep=0pt, node distance=0]
		
		\node[obs, label=260:$y_{k-N_p}$] at (0,-0.25) (y0){};
		\node[obs, label=270:$y_{k-N_p+1}$] at (1,-0.25) (y1){};
		\node at (2,-0.25) {$\dots$};
		\node[obs, label=270:$y_k$] at (3,-0.25) (yk){};
		\node at (4,-0.25) {$\dots$};
		\node[obs, label=270:$y_{k+N_d-2}$] at (5,-0.25) (ykp1){};
		\node[obs, label=280:$y_{k+N_d}$] at (6,-0.25) (yend){};
		
		\node[factor] at (3, 0.75) (qequ){};
		
		\node[latent, label=270:$x_{k-N_p'}$] at (0,1.75) (x1){};
		\node at (1.5,1.75) {$\dots$};
		\node[latent, label=305:$x_k$] at (3,1.75) (xk){};
		\node at (4.5,1.75) {$\dots$};
		\node[latent, label=270:$x_{k+N_d}$] at (6,1.75) (xend){};
		
		\node at (0,2.5) (qdem1){};
		\node[factor] at (3,2.75) (qdemk){};
		\node at (6,2.5) (qdemend){};

		\node at (0,4) {$\dots$};
		\node[latent, label=265:$d_{k,0}$] at (2,4) (ckq){};
		\node at (3,4) (cqkdots) {$\dots$};
		\node[latent] at (4,4) (cqkp1){};
		\node[latent, label=275:$d_{k,q-1}$] at (4,4) (cqkp1){};
		\node at (6,4) {$\dots$};

		\node[factor] at (2,5) (qdec_kq){};
		\node at (3,5) (qdecdots) {$\dots$};
		\node[factor] at (4,5) (qdec_kqp1){};

		\draw (1.5,0.5) [dashed, rounded corners=2mm]rectangle(4.5,1);
		\node[label=right:$f_\text{EQU}(\mathbf{x}_k)$] at (4.5, 0.75) (qequt){};
		\draw (2,2.5) [dashed, rounded corners=2mm]rectangle(4,3);
		\node[label=right:$f_\text{DEM}(x_k\text{, }\mathbf{d}_k)$] at (4, 2.75) (qdemt){};
		\draw (1.5,4.75) [dashed, rounded corners=2mm]rectangle(4.5,5.25);
		\node[label=right:$f_\text{DEC}(\mathbf{d}_k)$] at (4.5, 5) (qdect){};

		\draw [-] (y0) -- (qequ);
		\draw [-] (y1) -- (qequ);
		\draw [-] (yk) -- (qequ);
		\draw [-] (ykp1) -- (qequ);
		\draw [-] (yend) -- (qequ);
		
		\draw [-] (qequ) -- (x1);
		\draw [-] (qequ) -- (xk);
		\draw [-] (qequ) -- (xend);
		
		\draw [-] (ckq) -- (qdemk);
		\draw [-] (cqkp1) -- (qdemk);
		
		\draw [-, dotted] (qdem1) -- (x1);
		\draw [-] (qdemk) -- (xk);
		\draw [-, dotted] (qdemend) -- (xend);
		
		\draw (ckq) -- (qdec_kq);
		\draw (cqkp1) -- (qdec_kqp1);
		\end{tikzpicture}
		\caption{Factor graph for the posterior PDF  (\ref{eq_postfir}) on $x_k$ and $\mathbf{d}_k$.}
		\label{fig_fg_rec}
	\end{figure}

	A message-passing based decoding algorithm iteratively estimates the variable nodes (VN) $x_k$ and $d_{k,j}$ by using constraints imposed by factor nodes (FN). 
	Factor nodes are non proper PDFs for resolving transmission steps. The decoder FN models BICM encoding constraints with
	\begin{equation}
		f_{\text{DEC}}(d_{k,j}) \triangleq \textstyle p(d_{k,j}),
	\end{equation}
	and the demapper FN incorporates mapping constraints with 
	\begin{equation}
		f_{\text{DEM}}(x_k, \mathbf{d}_{k}) \triangleq \textstyle p(x_k\vert \mathbf{d}_k) = \prod_{j=0}^{q-1}\delta(d_{k,j}-\varphi_j^{-1}(x_k)), \label{eq_dem_fn}
	\end{equation}
	where $\delta$ is the Dirac delta function. The multipath channel constraints are modelled within the equalization factor node
	\begin{equation}
		f_{\text{EQU}}(\mathbf{x}_k) \triangleq  p(\mathbf{y}_k\vert\mathbf{x}_k) \propto e^{-\mathbf{y}_k^H\mathbf{y}_k/\sigma_w^2+2\mathcal{R}(\mathbf{y}_k^H\mathbf{H}_k\mathbf{x}_k)/\sigma_w^2}, \label{eq_equ_fn}
	\end{equation}
	where the dependence on $\mathbf{y}_k$ is omitted, as observations are fixed during the message-passing procedure. 
	Using these notations, the posterior (\ref{eq_postfir}) 
	gives the factor graph shown in Fig. \ref{fig_fg_rec}.

	\subsection{Expectation Propagation Message Passing Framework}\label{ssec:ep_framework}
	
	EP-based message passing algorithm is an extension of loopy belief propagation, where VNs are assumed to lie in the exponential distribution family \cite{minka2005divergence}. 
	Consequently, the exchanged messages are depicted by tractable distributions, and they allow iterative computation of a fully-factorized approximation for cumbersome posterior PDFs such as $p(\mathbf{\dbar}_k, \mathbf{x}_k \vert \mathbf{y}_k)$. 
	
	Updates at a FN  $\text{F}$  connected to variable nodes $\mathbf{v}$ are as follows.
	Messages exchanged between a VN $v_i$, the $i^\text{th}$ component of $\mathbf{v}$, and factor node $\text{F}$ are
	\begin{eqnarray}
		m_{v\rightarrow \text{F}}(v_i) &\triangleq&  \textstyle\prod_{\text{G} \neq F} m_{\text{G} \rightarrow v}(v_i), \label{eq_MESSfromVN} \\
		m_{\text{F}\rightarrow v}(v_i) &\triangleq& {\text{proj}_{\mathcal{Q}_{v_i}} \left[q_\text{F}(v_i)\right]}/{m_{v\rightarrow \text{F}}(v_i)},\label{eq_MESSfromFN}
	\end{eqnarray}
	where $\text{proj}_{\mathcal{Q}_{v_i}}$ is the Kullback-Leibler projection towards the probability distribution $\mathcal{Q}_{v_i}$ of VN $v_i$. The posterior $q_\text{F}(v_i)$ is an approximation of the marginal of the true posterior $p(\mathbf{v})$ on $v_i$, obtained by combining the true factor on FN F with messages from the neighbouring VNs
	\begin{equation}
		q_\text{F}(v_i) \triangleq \textstyle\int_{\mathbf{v}^{\backslash i}} f_\text{F}(\mathbf{v})\prod_{v_j} m_{v \rightarrow \text{F}}(v_j) \mathbf{dv}^{\backslash i},  \label{eq_postFN}
	\end{equation}
	where $\mathbf{v}^{\backslash i}$ are VNs without $v_i$ \cite{minka2005divergence}. 
	The projection operation for exponential families is equivalent to \emph{moment matching}, which simplifies the computation of messages \cite{minka_2001_expectationpropagation, minka2005divergence}.

	In this paper symbol VNs are assumed to lie in the family of multivariate circularly symmetric Gaussians with diagonal covariance matrices, making the approximate distributions fully factorized to independent Gaussians. Hence, a message on $x_k$ will be defined by a mean and a variance. The VNs $d_{k,j}$ are considered to follow Bernoulli distributions (which is included in the exponential family), and their messages can be described by bit log-likelihood ratios (LLR).

	This formalism is very generic and allows the derivation of many receiver structures.
	It has been used to derive a MIMO detector in \cite{senstAscheid_2011_frameworkEP_MMSEMIMO}, and a Kalman smoother in \cite{sunFleury_2015_iterativeRecvCombinedBPEP}. However EP receivers can also be derived without a message passing formalism, as recently shown for the block \cite{santosMurilloFuentes_2017_EPBLE} or FIR \cite{santosMurilloFuentes_2017_EPnuBLE} equalizers. To the authors' knowledge,  message-passing formalism was not previously used for FIR design, and it is favoured in this paper because of the available scheduling options it allows to clearly identify.

	\subsection{Derivation of Exchanged Messages}\label{ssec:ep_mess}

	\ifdouble
	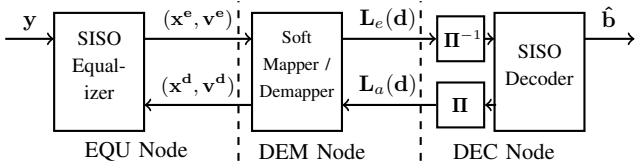
\begin{figure}[!t]
		\centering
		\begin{tikzpicture}[thick, scale=0.85, every node/.style={scale=0.85}]
		
		\draw [-, dashed] (1.7,-1.5) -- (1.7,1);
		\draw [-, dashed] (-1.15,-1.5) -- (-1.15,1);
		\node at (-2.75,-1.35) () {EQU Node};
		\node at (0,-1.35) () {DEM Node};
		\node at (3,-1.35) () {DEC Node};
		
		\matrix [ row sep = 0.05cm, column sep = 0.15cm, cells={scale=1.0}]
		{
			\node (ch_input)[xshift=-0cm] {}; &
			\node (eq_input)[yshift = 0.25cm, xshift=-0cm] {$\mathbf{y}$};
			&
			\node (dummy_equ_in1) [coordinate, xshift=-0.15cm] {};
			\node (dummy_equ1) [xshift=-0cm, yshift=0.3cm] {};&&&&&
			\node (dummy_equ_out1) [coordinate, xshift=0.15cm] {};
			&
			\node (eqoutput)[above]{\small $(\mathbf{{x}^e},\mathbf{v^e})$}; 
			&
			\node (dummy_dem_in1) [coordinate, xshift=-0.15cm] {};
			\node (dummy_dem1) [xshift=-0cm, yshift=0.3cm] {};&&&&&
			\node (dummy_dem_out1) [coordinate, xshift=0.15cm] {};
			&
			\node (llrx)[above]{$\mathbf{L}_e(\mathbf{d})$}; &
			\node (deintrlv) [block, fill=white] {\parbox[c]{.5cm}{\small \centering $\pmb{\Pi}^{-1}$}}; 
			&
			\node (dummy_dec_in1) [coordinate, xshift=-0.15cm] {};
			\node (dummy_dec1) [xshift=-0cm, yshift=0.3cm] {};&&&&&
			\node (dummy_dec_out1) [coordinate, xshift=0.15cm] {};
			&
			\node (dec_out) [yshift= 0.35cm]{$\mathbf{\hat{b}}$};
			&
			\node (dec_output)[coordinate] {};
			\\
			&&
			&&&&&
			\node (dummy_equ_in2) [coordinate, xshift=0.15cm] {};
			\node (dummy_equ2) [xshift=0cm, yshift=-0.3cm] {};
			&
			\node (x_mean)[above] {\small $(\mathbf{{x}^d},\mathbf{v^d})$};
			&
			\node (dummy_dem_out2) [coordinate, xshift=-0.15cm] {};
			&&&&&
			\node (dummy_dem_in2) [coordinate, xshift=0.15cm] {};
			\node (dummy_dem2) [xshift=0cm, yshift=-0.3cm] {};
			&
			\node (llrxe)[above]{$\mathbf{L}_a(\mathbf{d})$}; &
			\node (intrlv) [block, fill=white] {\parbox[c]{.5cm}{\small \centering $\pmb{\Pi}$}}; 
			&
			\node (dummy_dec_out2) [coordinate, xshift=-0.12cm] {};
			&&&&&
			\node (dummy_dec_in2) [coordinate, xshift=0.12cm] {};
			\node (dummy_dec2) [xshift=0cm, yshift=-0.3cm] {};
			&&
			\\
		};
		\node (equalizer) [fit=(dummy_equ1)(dummy_equ2), block] {\centering \small  SISO Equalizer};
		\node (demap) [fit=(dummy_dem1)(dummy_dem2), block] {\centering \footnotesize Soft Mapper / Demapper};
		\node (decode) [fit=(dummy_dec1)(dummy_dec2), block] {\centering \small SISO Decoder};
		
		\draw [->] (ch_input) -- (dummy_equ_in1);
		\draw [->] (dummy_equ_out1) -- (dummy_dem_in1);
		\draw [->] (dummy_dem_out1) -- (deintrlv);
		\draw [->] (deintrlv) -- (dummy_dec_in1);
		\draw [->] (dummy_dec_out1) -- (dec_output);
		\draw [->] (dummy_dec_out2) -- (intrlv);
		\draw [->] (intrlv) -- (dummy_dem_in2);
		\draw [->] (dummy_dem_out2) -- (dummy_equ_in2);
		
		\end{tikzpicture}
		\caption{Factor nodes shown as an iterative BICM receiver.}
		\label{fig_sim}
	\end{figure}
	\else
	\begin{figure}[!t]
		\centering
		\begin{tikzpicture}[thick, scale=0.85, every node/.style={scale=0.85}]
		
		\draw [-, dashed] (2.2,-2) -- (2.2,1.5);
		\draw [-, dashed] (-1.45,-2) -- (-1.45,1.5);
		\node at (-3.25,-1.75) () {EQU Node};
		\node at (0,-1.75) () {DEM Node};
		\node at (3.75,-1.75) () {DEC Node};
		
		\matrix [ row sep = 0.05cm, column sep = 0.2cm, cells={scale=1.0}]
		{
			\node (ch_input)[xshift=-0cm] {}; &
			\node (eq_input)[yshift = 0.25cm, xshift=-0cm] {$\mathbf{y}$};
			&
			\node (dummy_equ_in1) [coordinate, xshift=-0.15cm] {};
			\node (dummy_equ1) [xshift=-0cm, yshift=0.3cm] {};&&&&&
			\node (dummy_equ_out1) [coordinate, xshift=0.15cm] {};
			&
			\node (eqoutput)[above]{\small $(\mathbf{{x}^e},\mathbf{v^e})$}; 
			&
			\node (dummy_dem_in1) [coordinate, xshift=-0.15cm] {};
			\node (dummy_dem1) [xshift=-0cm, yshift=0.3cm] {};&&&&&
			\node (dummy_dem_out1) [coordinate, xshift=0.15cm] {};
			&
			\node (llrx)[above]{$\mathbf{L}_e(\mathbf{d})$}; &
			\node (deintrlv) [block, fill=white] {\parbox[c]{.5cm}{\small \centering $\pmb{\Pi}^{-1}$}}; 
			&
			\node (dummy_dec_in1) [coordinate, xshift=-0.15cm] {};
			\node (dummy_dec1) [xshift=-0cm, yshift=0.3cm] {};&&&&&
			\node (dummy_dec_out1) [coordinate, xshift=0.15cm] {};
			&
			\node (dec_out) [yshift= 0.35cm]{$\mathbf{\hat{b}}$};
			&
			\node (dec_output)[coordinate] {};
			\\
			&&
			&&&&&
			\node (dummy_equ_in2) [coordinate, xshift=0.15cm] {};
			\node (dummy_equ2) [xshift=0cm, yshift=-0.3cm] {};
			&
			\node (x_mean)[above] {\small $(\mathbf{{x}^d},\mathbf{v^d})$};
			&
			\node (dummy_dem_out2) [coordinate, xshift=-0.15cm] {};
			&&&&&
			\node (dummy_dem_in2) [coordinate, xshift=0.15cm] {};
			\node (dummy_dem2) [xshift=0cm, yshift=-0.3cm] {};
			&
			\node (llrxe)[above]{$\mathbf{L}_a(\mathbf{d})$}; &
			\node (intrlv) [block, fill=white] {\parbox[c]{.5cm}{\small \centering $\pmb{\Pi}$}}; 
			&
			\node (dummy_dec_out2) [coordinate, xshift=-0.12cm] {};
			&&&&&
			\node (dummy_dec_in2) [coordinate, xshift=0.12cm] {};
			\node (dummy_dec2) [xshift=0cm, yshift=-0.3cm] {};
			&&
			\\
		};
		\node (equalizer) [fit=(dummy_equ1)(dummy_equ2), block] {\centering \small  SISO Equalizer};
		\node (demap) [fit=(dummy_dem1)(dummy_dem2), block] {\centering \footnotesize Soft Mapper / Demapper};
		\node (decode) [fit=(dummy_dec1)(dummy_dec2), block] {\centering \small SISO Decoder};
		
		\draw [->] (ch_input) -- (dummy_equ_in1);
		\draw [->] (dummy_equ_out1) -- (dummy_dem_in1);
		\draw [->] (dummy_dem_out1) -- (deintrlv);
		\draw [->] (deintrlv) -- (dummy_dec_in1);
		\draw [->] (dummy_dec_out1) -- (dec_output);
		\draw [->] (dummy_dec_out2) -- (intrlv);
		\draw [->] (intrlv) -- (dummy_dem_in2);
		\draw [->] (dummy_dem_out2) -- (dummy_equ_in2);
		
		\end{tikzpicture}
		\caption{Factor nodes shown as an iterative BICM receiver.}
		\label{fig_sim}
	\end{figure}
	\fi
	
	This section details the EP-based message passing algorithm's application to the considered factor graph. First, exchanged messages are defined, and then their characterizing parameters are explicitly computed. See Fig.~\ref{fig_sim} for a conventional view of the receiver with these quantities.

	The messages arriving on the VN $x_k$ are Gaussians with
	\begin{eqnarray}
		m_{\text{EQU}\rightarrow x}(x_{k}) \propto \mathcal{CN}\left( x_k^e, v_k^e \right), \label{eq_equ_to_x} \\
		m_{\text{DEM}\rightarrow x}(x_{k}) \propto \mathcal{CN}\left( x_k^d, v_k^d\right), \label{eq_dem_to_x}
	\end{eqnarray}
	whereas messages arriving on the VN $d_{k,j}$ are Bernoullis
	\begin{eqnarray}
		m_{\text{DEC}\rightarrow d}(d_{k,j}) \propto \mathcal{B}\left( p_d^a \right),\,
		m_{\text{DEM}\rightarrow d}(d_{k,j}) \propto \mathcal{B}\left( p_d^e \right). \label{eq_FN_to_d}
	\end{eqnarray}
	During the message passing procedure, the characteristic parameters of these distributions are updated following a selected schedule. 
	For Bernoulli distributions, it is rather preferable to work with bit LLRs, rather than the success probability $p_d$:
	\begin{equation}
		L(d_j) \triangleq \ln\frac{\mathbb{P}[d_j=0]}{\mathbb{P}[d_j=1]} = \ln\frac{1-p_d}{p_d}.
	\end{equation}
	We use $L_a(\cdot)$, $L_e(\cdot)$ and $L(\cdot)$ operators to denote respectively a priori, extrinsic and a posteriori LLRs. When applied to $d_{k,j}$, this vocabulary represents the receiver's perspective, i.e. $L_a(d_{k,j})$, $L_e(d_{k,j})$ respectively characterize $m_{\text{DEC}\rightarrow d}(d_{k,j})$ and $m_{\text{DEM}\rightarrow d}(d_{k,j})$. 
	
	Finally, considering the factor graph shown on Fig. \ref{fig_fg_rec}, all variable nodes are only connected to a pair of distinct factor nodes. Consequently, using eq. (\ref{eq_MESSfromVN}), 
	$m_{v \rightarrow \text{F}}(v_i) = m_{\text{G}\rightarrow v}(v_i)$, 
	for all VN $v_i$, and FN $\text{F}, \text{G},\, \text{F}\neq\text{G}$ they are connected to.

	\subsubsection{Messages from DEC to DEM}
	
	In this paper, we assume DEC is a SISO decoder providing prior information $L_a(\mathbf{d})$ to DEM, whenever it receives extrinsic information $L_e(\mathbf{d})$ by DEM. 
	
	The demapper uses these prior LLRs, along with the DEM FN (\ref{eq_dem_fn}) to compute a prior PMF on $x_k=\alpha$, $\forall \alpha\in\mathcal{X}$ with
	\begin{equation}
		\mathcal{P}_{k}(\alpha) \propto \textstyle\prod_{j=0}^{q-1}e^{-\varphi^{-1}_j(\alpha)L_a(d_{k,j})}. \label{eq_priors}
	\end{equation}
	This is a categorical PMF corresponding to the marginal of $f_\text{DEM}(x_{k}, \mathbf{d}_{k})m_{d\rightarrow \text{DEC}}(\mathbf{d}_{k})$ on $x_{k}$ \cite{senstAscheid_2011_frameworkEP_MMSEMIMO}, used hereafter to compute approximate marginals $q_\text{DEM}(x_k)$ and $q_\text{DEM}(d_{k,j})$.
	
	\subsubsection{Messages from DEM to EQU}
	The demapper computes an approximate posterior on the VN $x_k$ using eq. (\ref{eq_postFN}) with
	\ifdouble
	\begin{equation}
		\begin{split}
			q_\text{DEM}(x_k) = \textstyle  \sum_{\mathbf{d}_k} & \textstyle  f_\text{DEM}(x_k, \mathbf{d}_k)
			m_{x \rightarrow \text{DEM}}(x_{k})\\
			& \textstyle \prod_{j=0}^{q-1} m_{d \rightarrow \text{DEM}}(d_{k,j}).
		\end{split}
	\end{equation}
	\else
	\begin{equation}
		q_\text{DEM}(x_k) = \textstyle   \sum_{\mathbf{d}_k} f_\text{DEM}(x_k, \mathbf{d}_k) m_{x \rightarrow \text{DEM}}(x_{k})
		\prod_{j=0}^{q-1} m_{d \rightarrow \text{DEM}}(d_{k,j}).
	\end{equation}
	\fi
	This is a posterior categorical PMF on the elements $x_k$ of $\mathcal{X}$, which can be computed using eqs. (\ref{eq_equ_to_x})
	and  (\ref{eq_priors}), which will be denoted as
	\begin{equation}
		\mathcal{D}_{k}(\alpha) \propto \exp{\left(-k_w{|\alpha-x^e_{k}|^2}/{v_k^e}\right)}\mathcal{P}_{k}(\alpha),\, \forall \alpha\in\mathcal{X}. \label{eq_posteriors}
	\end{equation}
	For computing messages towards EQU, the posterior PMF is projected into $\mathcal{CN}$ through moment matching. The mean and the variance of $\mathcal{D}_{k}$ are
	\begin{equation}
		\label{eq_dem_stats}
		\begin{aligned}
			\mu^d_k&\triangleq\mathbb{E}_{\mathcal{D}_k}[x_k]&={}&\textstyle\sum_{\alpha\in\mathcal{X}}\alpha \mathcal{D}_k(\alpha),\\
			\gamma^d_{k}&\triangleq\text{Var}_{\mathcal{D}_k}[x_k]&={}&\textstyle\sum_{\alpha\in\mathcal{X}}\vert\alpha\vert^2 \mathcal{D}_k(\alpha) - \vert \mu^d_k \vert^2.
		\end{aligned}
	\end{equation}
	When $m_{x \rightarrow \text{DEM}}(x_{k})\propto 1$, i.e. when there is no information from the EQU node (equivalent to $x_k^e=0$ and $v_k^e=+\infty$), $\mathcal{D}_{k}=\mathcal{P}_{k}$, and we denote the prior mean and variances as
	\begin{equation}
		{x}^p_k {}\triangleq{} \mathbb{E}_{\mathcal{P}_k}[x_k],\quad v^p_{k} {}\triangleq{} \text{Var}_{\mathcal{P}_k}[x_k].     \label{eq_prior_stats}
	\end{equation}
	Note that these values are used as soft feedback in conventional turbo equalization.
	
	Then in order to calculate $m_{\text{DEM}\rightarrow x}(x_{k})$ as in (\ref{eq_MESSfromFN}), a Gaussian division \cite{minka_2001_expectationpropagation} is implemented
	\begin{equation}
		{x}^*_k = \frac{\mu^d_k v^e_k - x^e_k\gamma^d_k} {v^e_k-\gamma^d_k},\, \text{and},\, {v}^*_k = \frac{ v^e_k \gamma^d_k} {v^e_k-\gamma^d_k}. \label{eq_demap_ep_new}
	\end{equation}
	This is the major novelty in using EP: the computation of an extrinsic feedback from the demapper to the equalizer.
	Attempting this with categorical distributions, as in BP, would completely remove $m_{x \rightarrow \text{DEM}}(x_{k})$, and the extrinsic ``feedback" to EQU would simply be the prior PMF $\mathcal{P}_k$ \cite{senstAscheid_2011_frameworkEP_MMSEMIMO}, which would yield a receiver equivalent to LE-IC \cite{tuchler_turbo_2002}.
	
	EP message passing algorithm consists in minimizing global divergence through iterative minimization of simpler local divergences. Thus, it might lock on undesirable fixed points, and a damping heuristic, as recommended in \cite[eq. (17)]{minka2005divergence}, is used to improve accuracy 
	\begin{equation}
		\label{eq_demap_ep_damp_stats_new}
		\begin{aligned}
			v_k^{d(\text{next})} &=  \left[(1-\beta)/v_k^{*} + \beta /\bar{v}_k^{d(\text{prev})}\right]^{-1}, \\
			x_{k}^{d(\text{next})} &=  {v}_k^{d(\text{next})}\left[(1-\beta)\frac{{x}_{k}^{*}}{v_k^{*}} + \beta \frac{{x}_{k}^{d(\text{prev})}}{{v}_k^{d(\text{prev})}}\right],
		\end{aligned}
	\end{equation}
	where $0 \leq \beta \leq 1$ configures the damping, and its effectiveness has been verified in \cite{santosMurilloFuentes_2017_EPnuBLE}.

	\subsubsection{Messages from EQU to DEM}
	The equalizer computes an approximate posterior on the VN $x_k$ using eq. (\ref{eq_postFN}) with
	\ifdouble
	\begin{equation}
		\begin{split}
			q_\text{EQU}(x_k) = \textstyle   \int_{\mathbf{x}^{\backslash k}_k} & \textstyle f_\text{EQU}(\mathbf{x}_k) \\
			& \textstyle \prod_{k'=k-N_p'}^{k+N_d}m_{x \rightarrow \text{EQU}}(x_{k'}) \mathbf{dx}^{\backslash k}_k.
		\end{split}
	\end{equation}
	\else
	\begin{equation}
		q_\text{EQU}(x_k) = \textstyle   \int_{\mathbf{x}^{\backslash k}_k} f_\text{EQU}(\mathbf{x}_k) \prod_{k'=k-N_p-L+1}^{k+N_d}m_{x \rightarrow \text{EQU}}(x_{k'}) \mathbf{dx}^{\backslash k}_k.
	\end{equation}
	\fi
	The integrand of the equation above is a multivariate Gaussian distribution $\mathcal{CN}(\pmb{\mu}^\mathbf{e}, \mathbf{\Gamma}^\mathbf{e})$, hence, using eq. (\ref{eq_equ_fn}), its covariance and mean satisfy
	\begin{equation}
		\begin{aligned}
			\mathbf{\Gamma}^\mathbf{e}_k &= (\mathbf{V}^\mathbf{d}_k{}^{-1}+\sigma_w^{-2}\mathbf{H}_k^H\mathbf{H}_k)^{-1}, \\
			\pmb{\mu}^\mathbf{e}_k &= \mathbf{\Gamma}^\mathbf{e}(\mathbf{V}^\mathbf{d}_k{}^{-1}\mathbf{{x}}^\mathbf{d}_k+\sigma_w^{-2}\mathbf{H}_k^H\mathbf{y}_k),
		\end{aligned}
	\end{equation}
	where $\mathbf{V}^\mathbf{d}_k=\textbf{diag}(\mathbf{v}^\mathbf{d}_k)$, with $\mathbf{v}^\mathbf{d}_k=[v^d_{k-N_p'},\dots,v^d_{k+N_d}]$, and $\mathbf{x}^\mathbf{d}_k=[x^d_{k-N_p'},\dots,x^d_{k+N_d}]$.
	Using some matrix algebra, and Woodbury's identity on $\mathbf{\Gamma}^\mathbf{e}$, the mean $\mu^e_k$ and the variance $\gamma^e_k$ of the marginalized PDF $q_\text{EQU}(x_k)$ are given by
	\begin{equation}
		\begin{aligned}
			\gamma_k^e &= \mathbf{e}_k^H{\mathbf{\Gamma}}^\mathbf{e}_k\mathbf{e}_k =v^d_k(1-v^d_k \mathbf{h}_k^H\mathbf{\Sigma}^\text{d}_k{}^{-1}\mathbf{h}_k), \\
			\mu_k^e &= \mathbf{e}_k^H{\pmb{\mu}}^\mathbf{e}_k = {x}^d_k + v^d_k\mathbf{h}_k^H\mathbf{\Sigma}^\text{d}_k{}^{-1}(\mathbf{y}_k-\mathbf{H}_k\mathbf{{x}}^\mathbf{d}_k),    
		\end{aligned}
	\end{equation}
	with $\mathbf{\Sigma}^\text{d}_k = k_w\sigma_w^2\mathbf{I}_N + \mathbf{H}_k\mathbf{V}^\mathbf{d}_k\mathbf{H}_k^H$.
	Message to the demapper is then extracted with the Gaussian density division in eq. (\ref{eq_MESSfromFN})
	\begin{equation}
		v^e_k = \frac{\gamma_k^e v^d_k}{v^d_k -\gamma_k^e},\, \text{and},\,
		{x}^e_k = \frac{v^d_k\mu_k^e -\gamma_k^e{x}^d_k}{v^d_k -\gamma_k^e}. \label{eq_equ_out}
	\end{equation}
	Developing these yields a FIR expression as in (\ref{eq_fir_model}) with $\mathbf{\bar{x}}_k^{\textbf{ep}}\triangleq [x^d_{k-N_p'},\dots,{x}^d_{k+N_d}]$ and $\mathbf{\bar{v}}_k^{\textbf{ep}} \triangleq [v_{k-N_p'}^{d},\dots,v_{k+N_d}^{d}]$ for IC.

	\subsubsection{Messages from DEM to DEC}
	The demapper computes an approximate posterior on the VN $d_{k,j}$ using eq. (\ref{eq_postFN}) with
	\ifdouble
	\begin{equation}
		\begin{split}
			q_\text{DEM}(\mathbf{d}_{k}) = \textstyle \sum_{x_k \in \mathcal{X}} 
			& \textstyle  f_\text{DEM}(x_k, \mathbf{d}_k) m_{x \rightarrow \text{DEM}}(x_{k}) \\
			& \textstyle \prod_{j=0}^{q-1} m_{d \rightarrow \text{DEM}}(d_{k,j}).
		\end{split}
	\end{equation}
	\else
	\begin{equation}
		q_\text{DEM}(\mathbf{d}_{k}) = \textstyle \sum_{x_k \in \mathcal{X}} f_\text{DEM}(x_k, \mathbf{d}_k) 
		m_{x \rightarrow \text{DEM}}(x_{k})
		\prod_{j=0}^{q-1} m_{d \rightarrow \text{DEM}}(d_{k,j}).
	\end{equation}
	\fi
	As bit LLRs are used to represent messages to DEC, this distribution is marginalized on $d_{k,0},\dots,d_{k,q-1}$ \cite{senstAscheid_2011_frameworkEP_MMSEMIMO}, and the division in eq. (\ref{eq_MESSfromFN}) is directly carried out with
	LLRs
	\begin{equation}
		L_e(d_{k,j}) = \ln \sum_{\alpha\in\mathcal{X}_j^0}{\mathcal{D}_k(\alpha)} - \ln \sum_{\alpha\in\mathcal{X}_j^1}{\mathcal{D}_k(\alpha)} - L_a(d_{k,j}), \label{eq_demap_llr}
	\end{equation}
	with $\mathcal{X}_j^p=\lbrace \alpha\in\mathcal{X}: \varphi^{-1}_j(x)=p\rbrace$ where $p\in\mathbb{F}_2$.

	\begin{algorithm}[!t]
		\caption{Proposed \emph{Self-Iterated DFE-IC EP} receiver.}
		\label{algo:fdsidfeicep}
		\begin{algorithmic}[1]
			\renewcommand{\algorithmicrequire}{\textbf{Input}}
			\renewcommand{\algorithmicensure}{\textbf{Output}}
			\REQUIRE $\mathbf{y}$, $\mathbf{H}$, $\sigma_w^2$
			\STATE Initialize decoder with $L_a^{(0)}\left(\mathbf{d}_{k}\right)=0, \forall k$.
			\FOR {$\tau = 0$ to $\mathcal{T}$}
			\STATE  $\forall k=0,\dots,K-1$, use $L_a^{(\tau)}(\mathbf{d})$ to compute $\mathcal{P}^{(\tau)}_k$ with (\ref{eq_priors}), and set $(x_k^{d(0)}, v_k^{d(0)})\gets (x_k^{p}, v_k^{p})$ using (\ref{eq_prior_stats}).
			\FOR {$s = 0$ to $\mathcal{S}_\tau$}
			\FOR {$k = 0$ to $K-1$}
			\STATE Equalize using (\ref{eq_equ_out}) and get $(x_k^{e(s)}, v_k^{e(s)})$.
			\STATE Use (\ref{eq_posteriors})-(\ref{eq_dem_stats}) to update $\mathcal{D}^{(s+1)}_{k}$, and generate EP feedback $(x_k^{d(s+1)}, v_k^{d(s+1)})$ with (\ref{eq_demap_ep_new})-(\ref{eq_demap_ep_damp_stats_new}). 
			\STATE If $v_k^{d(s+1)}\leq 0$, then $(x_k^{d(s+1)}, v_k^{d(s+1)}) \gets (\mu_k^d, \gamma_k^d)$ and store $k$ in the set $\mathcal{I}_\text{err}^{(s)}$.		
			\ENDFOR
			\STATE $\forall k\in \mathcal{I}^{(s)}_\text{err}$, $(x_k^{d(s+1)}, v_k^{d(s+1)}) \gets (x_k^{d(s)}, v_k^{d(s)})$.
			\ENDFOR
			\STATE Compute $L^{(\tau)}_e(\mathbf{d}_{k})$ using $\mathcal{D}^{(\tau, \mathcal{S}_\tau)}_{k}$ with (\ref{eq_demap_llr}), $\forall k$, and provide them to the decoder, to obtain $L^{(\tau+1)}_a(\mathbf{d}_k), \forall k$.
			\ENDFOR
		\end{algorithmic} 
	\end{algorithm}

	\subsection{Proposed Self-Iterated DFE-IC EP Receiver} \label{ssec:dfeic_ep}

	A factor graph (sec. \ref{ssec:ep_factorgraph}) and messages exchanged over it (sec. \ref{ssec:ep_mess}) are necessary to derive a receiver algorithm, but may be insufficient when considering a graph with cycles. 
	Indeed, specifying a scheduling for the update of VNs and FNs is also required.
	
	In this paper, a serial scheduling across variable nodes $x_k$ is considered. In detail, when EQU updates a VN $x_k$, factor node DEM is immediately activated in order to provide its own extrinsic estimation of $x_k$, jointly using prior information from the decoder and the equalizer's extrinsic output. This results in a DFE-IC structure, using a novel kind of soft feedback, unlike any hard or soft APP feedback previously used in the literature \cite{tuchler_turbo_2002,jeong_turbo_2010,jeong_soft-soft-out_2011,lopes_soft-feedback_2006,tao_2016_low,trajkovic_turbo_2005,balakrishnan_mitigation_1999,lou_soft-decision_2011}. Moreover, when detection across the whole block is completed, this serial scheduling can be repeated by keeping the previously updated DEM messages, yielding a \emph{self-iterated DFE-IC EP} structure. 
	
	To clarify the dynamics of the proposed receiver, $\tau=0, \dots, \mathcal{T}$ denotes turbo iterations (TI), i.e. exchanges between the DEM and DEC factor nodes. Each TI consists of $s=0, \dots, \mathcal{S}_\tau$ self-iterations (SI) (may vary with $\tau$), i.e. exchanges between EQU and DEM factor nodes, which  \emph{sequentially} updates the whole block $\mathbf{x}$. In the following, EQU $\leftrightarrow$ DEM messages derived previously are appended a superscript $(s)$, but $\tau$ is omitted for readability.

	The proposed scheduling, given in Algorithm \ref{algo:fdsidfeicep}, generates an EP FIR receiver which uses the following means and variances for interference cancellation
	\begin{equation}
		\label{eq_dfeep_stats}
		\begin{split}
			\mathbf{\bar{x}}^\textbf{dfe-ep}_k{}^{(s)} \triangleq [{x}^{d(s+1)}_{k-N_p'},\dots,{x}^{d(s+1)}_{k-1},{x}^{d(s)}_k,\dots, {x}^{d(s)}_{k+N_d}]^T,\\
			\mathbf{\bar{v}}^\textbf{dfe-ep}_k{}^{(s)} \triangleq [v^{d(s+1)}_{k-N_p'},\dots,v^{d(s+1)}_{k-1},v^{d(s)}_k,\dots, v^{d(s)}_{k+N_d}]^T, 
		\end{split}
	\end{equation}
	for $k = 0,\dots,K-1$. This layout shows that this structure indeed follows a time-varying DFE-IC evolution, with anti-causal symbols using demapper's output from the previous self-iteration, and causal symbols using current EP feedback from the demapper. The extrinsic feedback from the demapper is obtained by using jointly the prior information from the previous TI, and the past equalizer outputs of the current and previous self iterations (see (\ref{eq_posteriors})-(\ref{eq_demap_ep_damp_stats_new})). The Algorithm \ref{algo:fdsidfeicep} also incorporates a mechanism to deal with EP-based feedback's infamous negative variances \cite{senstAscheid_2011_frameworkEP_MMSEMIMO,santosMurilloFuentes_2017_EPBLE}, with the set $\mathcal{I}^{(s)}_\text{err}$ which stores their indexes. These values are replaced with APP-based variances in the current SI, and then replaced again with their previous values for the next SI.

	\ifdouble
	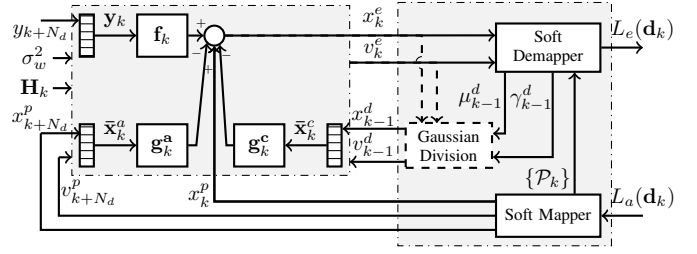
\begin{figure}[!t]
		\centering
		\begin{tikzpicture}[xscale=0.8, yscale=0.8, every node/.style={scale=0.8}, thick]
		\def\radius{1.mm} 
		\matrix [row sep = 0.15cm, column sep = 0.025cm, cells={scale=0.8}]
		{
			\node (dummy_chanest_in) [coordinate, xshift=-0.4cm] {}; 
			\node (dummy_obs_in) [coordinate, xshift=-0.6cm] {};		
			&
			\node (dummy_1) [coordinate, xshift=0cm, yshift = 0.0cm] {};
			\node (sw_y) [block, preaction={fill, white}, pattern=horizontal lines] {\parbox[c]{.2cm}{}};
			&
			\node[yshift=0.25cm]{$\mathbf{y}_k$};
			&
			\node (filter1) [block, fill = white, xshift=0.0cm] {\parbox[c]{0.6cm}{\centering $\mathbf{f}_k$}};
			&
			\node (dummy_sum_offset) [coordinate, xshift=0cm] {};
			\node (sum2) [sum, xshift=0.4cm] {};
			&
			\node (dummy_2) [coordinate, xshift=0cm] {};
			\node (dummy_sum) [coordinate] {};
			&&
			&
			\node (dummy_var_out) [coordinate, xshift=0.25cm, yshift=-0.2cm] {};
			\node (dummy_var_out2) [coordinate, xshift=0.25cm, yshift=-0.45cm] {};
			&
			\node[xshift=0cm,yshift=0.3cm](output){${{x}}^e_k$};
			\node[xshift=0cm,yshift=-0.215cm](var_out){$v^e_k$};
			&&
			\node (demap) [block, fill = white, yshift=-0.2cm] {\parbox[c]{1.5cm}{\centering \footnotesize Soft Demapper}};
			&
			\node (Le)[xshift=0cm, yshift=0.1cm]{$L_e(\mathbf{d}_k)$};
			\node (dummy_Le) [coordinate, xshift=0cm,yshift=-0.2cm] {};
			\\
			\vspace{2cm}
			\\
			\node (dummy_ap_in) [coordinate, xshift=-0.6cm, yshift=0.2cm] {}; 
			\node (dummy_v_ap_in) [coordinate, xshift=-0.3cm, yshift=-0.2cm] {}; 
			&
			\node (dummy_4) [coordinate, xshift=0cm] {};
			\node (dummy_v_ap_in2) [coordinate, xshift=-0.25cm, yshift=-0.2cm] {};
			\node (dummy_sw_ac) [coordinate, xshift=-0.1cm, yshift=0.2cm] {};
			\node (sw_ac) [block, , preaction={fill, white}, pattern=horizontal lines, xshift=0.cm] {\parbox[c]{.2cm}{}};
			&
			\node[yshift=0.25cm]{$\mathbf{\bar{x}}^a_k$};
			&
			\node (filter2) [block, fill = white, xshift=0cm] {\parbox[c]{0.6cm}{\centering $\mathbf{g}^\mathbf{a}_k$}}; 
			&
			\node (dummy_filt2sum) [coordinate, xshift=0.1cm] {};
			\node (dummy_filt3sum) [coordinate, xshift=0.6cm] {};
			&&
			\node (filter3) [block, fill = white, xshift=0.5cm] {\parbox[c]{0.6cm}{\centering $\mathbf{g}^\mathbf{c}_k$}}; 
			&
			\node[yshift=0.25cm]{$\mathbf{\bar{x}}^c_k$};
			&
			\node (sw_c) [block, , preaction={fill, white}, pattern=horizontal lines] {\parbox[c]{.2cm}{}};
			\node (dummy_3) [coordinate, xshift=0cm] {};
			&
			\node[yshift=0.5cm, xshift=0cm]{${{x}}^d_{k-1}$};
			\node[yshift=0cm, xshift=0cm]{${{v}}^d_{k-1}$};
			&
			\node (x_p_in)[xshift=-0.45cm, yshift=0.37cm, coordinate] {};
			\node (v_p_in)[xshift=-0.2cm, yshift=0.37cm, coordinate] {};
			\node (gaussdiv) [block, fill = white, dashed] {\parbox[c]{1.2cm}{\centering \footnotesize Gaussian Division}};
			&
			\\
			&&&
			&
			&&&&&&
			&
			\node (soft) [block, fill = white] {\parbox[c]{1.5cm}{\centering \footnotesize Soft Mapper}}; 
			&
			\node (La_up)[xshift=0cm, yshift=0.3cm]{$L_a(\mathbf{d}_k)$};	
			\node (dummy_ap) [coordinate, xshift=0cm, yshift = 0cm] {};
			\\
		};
		\begin{scope}[on background layer]	
		\node [coordinate, xshift=-2.5cm, yshift=0.75cm](nAux0) at (demap) {};
		\node [coordinate, xshift=1.05cm, yshift=-0.5cm](nAux1) at (soft) {};
		\draw [dashdotted, fill=gray, fill opacity=0.1] (nAux0) -| (nAux1) -| (nAux0) {};
		
		\node [coordinate, xshift=-0.25cm, yshift=0cm](nAux0) at (dummy_1) {};
		\node [coordinate, xshift=-0.25cm, yshift=0.5cm](nAux1) at (dummy_1) {};
		\node [coordinate, xshift=0cm, yshift=0.5cm](nAux2) at (dummy_2) {};
		\node [coordinate, xshift=0.25cm, yshift=-0.5cm](nAux3) at (dummy_3) {};
		\node [coordinate, xshift=0.25cm, yshift=0cm](nAux03) at (dummy_3) {};
		\node [coordinate, xshift=0cm, yshift=-0.5cm](nAux4) at (dummy_4) {};
		\draw [dashdotted, fill=gray, fill opacity=0.1] (nAux1) -| (nAux2) -| (nAux3) -| (nAux4) -| (nAux1) {};
		\end{scope}
		

		\draw [->, transform canvas={yshift=-0.3cm}] (dummy_chanest_in) -- node[pos=0,xshift=-0.3cm]{$\sigma_w^2$}(nAux0);
		\draw [->, transform canvas={yshift=-0.7cm}] (dummy_chanest_in) -- node[pos=0,xshift=-0.3cm]{$\mathbf{H}_k$}(nAux0);

		\draw [->, transform canvas={yshift=0.2cm}] (dummy_obs_in) -- node[pos=0,yshift=-0.2cm]{$y_{k+N_d}$}(sw_y);
		\draw [->] (sw_y) -- (filter1);
		\draw [->] (filter1) -- node[pos=0.85, yshift=0.20cm]{\tiny{$+$}} (sum2);
		\draw [->] (sw_ac) -- (filter2);
		\draw [->] (filter2.0) -- (dummy_filt2sum) -- node[pos=0.9, xshift=-0.15cm]{\tiny{$-$}}(sum2.225);
		\draw [->] (filter3.180) -- (dummy_filt3sum) -- node[pos=0.9, xshift=0.10cm]{\tiny{$-$}}(sum2.300);
		
		\draw [->] (dummy_ap) -- (soft);
		\draw [->] (demap) -- (dummy_Le);
		\draw [->,  name path = exout] (sum2) -- (dummy_sum) -- (demap.167);
		\draw [->, transform canvas={yshift=-0.2cm},  name path = evout] (dummy_var_out) -- (demap);
		
		\draw [-] (soft.195) -| node[pos=1,yshift=0.25cm]{$x^p_{k+N_d}$}(dummy_ap_in);
		\draw [->] (dummy_ap_in) -- (dummy_sw_ac);
		\draw [-] (soft.180) -| node[pos=0.7,xshift=0.5cm]{$v^p_{k+N_d}$}(dummy_v_ap_in);
		\draw [->] (dummy_v_ap_in) -- (dummy_v_ap_in2);
		\draw [->] (soft.165) -| node[pos=0.925, xshift=-0.1cm]{\tiny{$+$}}(sum2);
		\draw [->] (soft.165) -| node[pos=0.52, xshift=-0.25cm]{$x^p_{k}$}(sum2);

		\draw [->, transform canvas={yshift=0.22cm}] (gaussdiv) -- (sw_c);
		\draw [->, transform canvas={yshift=-0.22cm}] (gaussdiv) -- (nAux03);
		\draw [->, transform canvas={yshift=-0cm}] (sw_c) -- (filter3);
		\node [coordinate, xshift=0.65cm, yshift=0.6cm](nAux1) at (filter3) {};
		
		\draw [->, transform canvas={xshift=0.35cm}] (soft) -- node[pos=0.125, xshift=-0.45cm]{$\tiny \lbrace\mathcal{P}_k\rbrace$}(demap);
		
		\draw [->, transform canvas={xshift=0cm}] (demap.210) |- node[pos=0.2, xshift=-0.4cm]{$\mu^d_{k-1}$}(gaussdiv.15);	
		\draw [->, transform canvas={xshift=0cm}] (demap.280) |- node[pos=0.15, xshift=-0.35cm]{$\gamma^d_{k-1}$}(gaussdiv.345);	
		
		\draw[->, dashed] (dummy_var_out2) -| (v_p_in);
		\path [->,  name path = gdxin] (sum2) -| (x_p_in);
		\path [name intersections={of = evout and gdxin}];
		\coordinate[yshift=-0.25cm] (S)  at (intersection-1);
		\path[name path=circle] (S) circle(\radius);
		\path [name intersections={of = circle and gdxin}];
		\coordinate (I1)  at (intersection-1);
		\coordinate (I2)  at (intersection-2);
		\tkzDrawArc[color=black, dashed](S,I1)(I2);
		\draw[dashed] (sum2) -| (I1);
		\draw[->, dashed] (I2) -- (x_p_in);
		
		\end{tikzpicture}
		\caption{TV DFE-IC EP (dashed) / APP (no dashed) structure.}
		\label{fig_dfe_prop}
	\end{figure}
	\else
	\begin{figure}[!t]
		\centering
		\begin{tikzpicture}[xscale=0.8, yscale=0.8, every node/.style={scale=0.8}, thick]
		\def\radius{1.mm} 
		\matrix [row sep = 0.15cm, column sep = 0.025cm, cells={scale=0.8}]
		{
			\node (dummy_chanest_in) [coordinate, xshift=-0.4cm] {}; 
			\node (dummy_obs_in) [coordinate, xshift=-0.6cm] {};		
			&
			\node (dummy_1) [coordinate, xshift=0cm, yshift = 0.0cm] {};
			\node (sw_y) [block, preaction={fill, white}, pattern=horizontal lines] {\parbox[c]{.2cm}{}};
			&
			\node[yshift=0.25cm]{$\mathbf{y}_k$};
			&
			\node (filter1) [block, fill = white, xshift=0.0cm] {\parbox[c]{0.6cm}{\centering $\mathbf{f}_k$}};
			&
			\node (dummy_sum_offset) [coordinate, xshift=0cm] {};
			\node (sum2) [sum, xshift=0.4cm] {};
			&
			\node (dummy_2) [coordinate, xshift=0cm] {};
			\node (dummy_sum) [coordinate] {};
			&&
			&
			\node (dummy_var_out) [coordinate, xshift=0.25cm, yshift=-0.2cm] {};
			\node (dummy_var_out2) [coordinate, xshift=0.25cm, yshift=-0.45cm] {};
			&
			\node[xshift=0cm,yshift=0.3cm](output){${{x}}^e_k$};
			\node[xshift=0cm,yshift=-0.215cm](var_out){$v^e_k$};
			&&
			\node (demap) [block, fill = white, yshift=-0.2cm] {\parbox[c]{1.5cm}{\centering \footnotesize Soft Demapper}};
			&
			\node (Le)[xshift=0cm, yshift=0.1cm]{$L_e(\mathbf{d}_k)$};
			\node (dummy_Le) [coordinate, xshift=0cm,yshift=-0.2cm] {};
			\\
			\vspace{2cm}
			\\
			\node (dummy_ap_in) [coordinate, xshift=-0.6cm, yshift=0.2cm] {}; 
			\node (dummy_v_ap_in) [coordinate, xshift=-0.3cm, yshift=-0.2cm] {}; 
			&
			\node (dummy_4) [coordinate, xshift=0cm] {};
			\node (dummy_v_ap_in2) [coordinate, xshift=-0.25cm, yshift=-0.2cm] {};
			\node (dummy_sw_ac) [coordinate, xshift=-0.1cm, yshift=0.2cm] {};
			\node (sw_ac) [block, , preaction={fill, white}, pattern=horizontal lines, xshift=0.cm] {\parbox[c]{.2cm}{}};
			&
			\node[yshift=0.25cm]{$\mathbf{\bar{x}}^a_k$};
			&
			\node (filter2) [block, fill = white, xshift=0cm] {\parbox[c]{0.6cm}{\centering $\mathbf{g}^\mathbf{a}_k$}}; 
			&
			\node (dummy_filt2sum) [coordinate, xshift=0.1cm] {};
			\node (dummy_filt3sum) [coordinate, xshift=0.6cm] {};
			&&
			\node (filter3) [block, fill = white, xshift=0.5cm] {\parbox[c]{0.6cm}{\centering $\mathbf{g}^\mathbf{c}_k$}}; 
			&
			\node[yshift=0.25cm]{$\mathbf{\bar{x}}^c_k$};
			&
			\node (sw_c) [block, , preaction={fill, white}, pattern=horizontal lines] {\parbox[c]{.2cm}{}};
			\node (dummy_3) [coordinate, xshift=0cm] {};
			&
			\node[yshift=0.55cm, xshift=0cm]{${{x}}^d_{k-1}$};
			\node[yshift=0cm, xshift=0cm]{${{v}}^d_{k-1}$};
			&
			\node (x_p_in)[xshift=-0.45cm, yshift=0.53cm, coordinate] {};
			\node (v_p_in)[xshift=-0.2cm, yshift=0.53cm, coordinate] {};
			\node (gaussdiv) [block, fill = white, dashed] {\parbox[c]{1.2cm}{\centering \footnotesize Gaussian Division}};
			&
			\\
			&&&
			&
			&&&&&&
			&
			\node (soft) [block, fill = white] {\parbox[c]{1.5cm}{\centering \footnotesize Soft Mapper}}; 
			&
			\node (La_up)[xshift=0cm, yshift=0.3cm]{$L_a(\mathbf{d}_k)$};	
			\node (dummy_ap) [coordinate, xshift=0cm, yshift = 0cm] {};
			\\
		};
		\begin{scope}[on background layer]	
		\node [coordinate, xshift=-2.5cm, yshift=0.75cm](nAux0) at (demap) {};
		\node [coordinate, xshift=1.05cm, yshift=-0.5cm](nAux1) at (soft) {};
		\draw [dashdotted, fill=gray, fill opacity=0.1] (nAux0) -| (nAux1) -| (nAux0) {};
		
		\node [coordinate, xshift=-0.25cm, yshift=0cm](nAux0) at (dummy_1) {};
		\node [coordinate, xshift=-0.25cm, yshift=0.5cm](nAux1) at (dummy_1) {};
		\node [coordinate, xshift=0cm, yshift=0.5cm](nAux2) at (dummy_2) {};
		\node [coordinate, xshift=0.25cm, yshift=-0.5cm](nAux3) at (dummy_3) {};
		\node [coordinate, xshift=0.25cm, yshift=0cm](nAux03) at (dummy_3) {};
		\node [coordinate, xshift=0cm, yshift=-0.5cm](nAux4) at (dummy_4) {};
		\draw [dashdotted, fill=gray, fill opacity=0.1] (nAux1) -| (nAux2) -| (nAux3) -| (nAux4) -| (nAux1) {};
		\end{scope}
		

		\draw [->, transform canvas={yshift=-0.3cm}] (dummy_chanest_in) -- node[pos=0,xshift=-0.3cm]{$\sigma_w^2$}(nAux0);
		\draw [->, transform canvas={yshift=-0.7cm}] (dummy_chanest_in) -- node[pos=0,xshift=-0.3cm]{$\mathbf{H}_k$}(nAux0);

		\draw [->, transform canvas={yshift=0.2cm}] (dummy_obs_in) -- node[pos=0,yshift=-0.2cm]{$y_{k+N_d}$}(sw_y);
		\draw [->] (sw_y) -- (filter1);
		\draw [->] (filter1) -- node[pos=0.85, yshift=0.20cm]{\tiny{$+$}} (sum2);
		\draw [->] (sw_ac) -- (filter2);
		\draw [->] (filter2.0) -- (dummy_filt2sum) -- node[pos=0.9, xshift=-0.15cm]{\tiny{$-$}}(sum2.225);
		\draw [->] (filter3.180) -- (dummy_filt3sum) -- node[pos=0.9, xshift=0.10cm]{\tiny{$-$}}(sum2.300);
		
		\draw [->] (dummy_ap) -- (soft);
		\draw [->] (demap) -- (dummy_Le);
		\draw [->,  name path = exout] (sum2) -- (dummy_sum) -- (demap.167);
		\draw [->, transform canvas={yshift=-0.2cm},  name path = evout] (dummy_var_out) -- (demap);
		
		\draw [-] (soft.195) -| node[pos=1,yshift=0.3cm]{$x^p_{k+N_d}$}(dummy_ap_in);
		\draw [->] (dummy_ap_in) -- (dummy_sw_ac);
		\draw [-] (soft.180) -| node[pos=0.7,xshift=0.6cm]{$v^p_{k+N_d}$}(dummy_v_ap_in);
		\draw [->] (dummy_v_ap_in) -- (dummy_v_ap_in2);
		\draw [->] (soft.165) -| node[pos=0.925, xshift=-0.1cm]{\tiny{$+$}}(sum2);
		\draw [->] (soft.165) -| node[pos=0.54, xshift=-0.3cm]{$x^p_{k}$}(sum2);

		\draw [->, transform canvas={yshift=0.22cm}] (gaussdiv) -- (sw_c);
		\draw [->, transform canvas={yshift=-0.22cm}] (gaussdiv) -- (nAux03);
		\draw [->, transform canvas={yshift=-0cm}] (sw_c) -- (filter3);
		\node [coordinate, xshift=0.65cm, yshift=0.6cm](nAux1) at (filter3) {};
		
		\draw [->, transform canvas={xshift=0.35cm}] (soft) -- node[pos=0.125, xshift=-0.45cm]{$\tiny \lbrace\mathcal{P}_k\rbrace$}(demap);
		
		\draw [->, transform canvas={xshift=0cm}] (demap.220) |- node[pos=0.2, xshift=-0.4cm]{$\mu^d_{k-1}$}(gaussdiv.15);	
		\draw [->, transform canvas={xshift=0cm}] (demap.280) |- node[pos=0.15, xshift=-0.375cm]{$\gamma^d_{k-1}$}(gaussdiv.345);	
		
		\draw[->, dashed] (dummy_var_out2) -| (v_p_in);
		\path [->,  name path = gdxin] (sum2) -| (x_p_in);
		\path [name intersections={of = evout and gdxin}];
		\coordinate[yshift=-0.25cm] (S)  at (intersection-1);
		\path[name path=circle] (S) circle(\radius);
		\path [name intersections={of = circle and gdxin}];
		\coordinate (I1)  at (intersection-1);
		\coordinate (I2)  at (intersection-2);
		\tkzDrawArc[color=black, dashed](S,I1)(I2);
		\draw[dashed] (sum2) -| (I1);
		\draw[->, dashed] (I2) -- (x_p_in);
		
		\end{tikzpicture}
		\caption{MMSE DFE-IC EP(dashed)/APP(no dashed) structure.}
		\label{fig_dfe_prop}
	\end{figure}
	\fi

	Although equation (\ref{eq_fir_model}) is useful for FIR analysis, causal and anti-causal feedback of DFE-IC should be separated in practice. Using
	\begin{eqnarray}
		\mathbf{E}^\textbf{c} = [\mathbf{I}_{N_p'},\, \mathbf{0}_{N_p', N_d+1}],\, 
		\mathbf{E}^\textbf{a} = [\mathbf{0}_{N_d+1, N_p'},\, \mathbf{I}_{N_d+1}],
	\end{eqnarray}
	we define
	$\mathbf{H}^\textbf{c}_k =  \mathbf{H}_k\mathbf{E}^\textbf{c}{}^T$ and
	$\mathbf{H}^\textbf{a}_k = \mathbf{H}_k\mathbf{E}^\textbf{a}{}^T$,
	to respectively operate on $\mathbf{\bar{x}}_k^{\textbf{c}(s)}=\mathbf{E}^\textbf{c}\mathbf{\bar{x}}^{\textbf{dfe-ep}(s)}_k$, and $\mathbf{\bar{x}}_k^{\textbf{a}(s)}=\mathbf{E}^\textbf{a}\mathbf{\bar{x}}^{\textbf{dfe-ep}(s)}_k$, as a generalized interference cancellation scheme. The \emph{SI DFE-IC EP} of (\ref{eq_dfeep_stats}), is rewritten as:
	\begin{equation}
		\label{eq_dfeic}
		\begin{aligned}
			{x}^{e(s)}_k &= \bar{x}^{a(s)}_k + \mathbf{f}_k^{(s)}{}^H\mathbf{y}_k - \mathbf{g}^{\mathbf{c}(s)}_k{}^H\mathbf{\bar{x}}_k^{\mathbf{c}(s)} - \mathbf{g}^{\mathbf{a}(s)}_k{}^H\mathbf{\bar{x}}_k^{\mathbf{a}(s)}, 
			\\
			v^{e(s)}_k &= 1/\xi^{\text{dfe-ep}(s)}_k  - \bar{v}^{a(s)}_k,
		\end{aligned} 
	\end{equation}
	with $\mathbf{f}_k^{(s)} =\mathbf{\Sigma}_k^{\textbf{dfe-ep}(s)}{}^{-1}\mathbf{h}_k/\xi^{\text{dfe-ep}(s)}_k$,
	$\mathbf{g}^{\textbf{c}(s)}_k = \mathbf{H}^\textbf{c}_k{}^H\mathbf{f}_k$, and 
	$\mathbf{g}^{\textbf{a}(s)}_k = \mathbf{H}^\textbf{a}_k{}^H\mathbf{f}^{(s)}_k$. When $\mathcal{S}_\tau=0$, the proposed receiver is a strict TV DFE-IC EP, with ${\cdot}^{d(s+1)}_{k}={\cdot}^{d}_{k}$ and ${\cdot}^{d(s)}_{k} = {\cdot}^{p}_{k}$, this case is shown on Fig. \ref{fig_dfe_prop} with the dashed module.

	In conclusion, we have applied message passing framework of EP for equalization, using \emph{sliding window} observations. This results in a novel message computation given by (\ref{eq_demap_ep_new})-(\ref{eq_demap_ep_damp_stats_new}) and (\ref{eq_equ_out}), unlike blockwise messages in \cite{senstAscheid_2011_frameworkEP_MMSEMIMO,santosMurilloFuentes_2017_EPBLE}. Moreover, by using an hybrid serial/parallel schedule, our structure operates as a self-iterated DFE-IC, unlike the self-iterated LE-IC scheme concurrently developed in \cite{santosMurilloFuentes_2017_EPnuBLE}.
	In the following, a matrix inversion strategy is introduced, that reduces the computational complexity difference between DFE-IC and LE-IC.

	\begin{algorithm}[t!]
		\caption{Cholesky update algorithm for LE-IC.}
		\label{algo:lecholupdate}
		\begin{algorithmic}[1]
			\renewcommand{\algorithmicrequire}{\textbf{Input}}
			\renewcommand{\algorithmicensure}{\textbf{Output}}
			\REQUIRE $\mathbf{L}_{k-1}, \sigma_w^2, \bar{v}_{k+N_d}, \mathbf{H}_{k-1}, \mathbf{H}_{k}, \mathbf{\bar{V}}_{k-1}$
			\ENSURE  $\mathbf{L}_{k}$
			\STATE \COMMENT{\textit{Add a row and a column}}
			\STATE $[\mathbf{h_1}_k,{h_2}_k] \gets {[0, \mathbf{e}_{k+N_d}^H\mathbf{H}_{k}]} $
			\STATE $\mathbf{w} \gets \mathbf{H}_{k-1}\mathbf{\bar{V}}_{k-1}\mathbf{h_1}_k $
			\STATE $\mathbf{l_{12}} \gets \mathbf{L}_{k-1}^{-1}\mathbf{w} $
			\STATE $ l_{22} \gets \sqrt{\mathbf{h_1}_k^H\mathbf{\bar{V}}_{k-1}\mathbf{h_1}_k+\bar{v}_{k+N_d}|{h_2}_k|^2-\mathbf{l_{12}}^H\mathbf{l_{12}}+\sigma_w^2}$
			\STATE \COMMENT{\textit{Build augmented matrix and remove row \& column}}
			\STATE $\left[\begin{array}{cc}
			\times & \mathbf{0}_{1, N}  \\ 
			\mathbf{l_{21}} & \mathbf{L_{22}}
			\end{array} \right] \gets \left[\begin{array}{cc}
			\mathbf{L}_{k-1} & \mathbf{0}_{N, 1}  \\ 
			\mathbf{l_{12}}^H & l_{22}
			\end{array} \right]$
			\STATE \COMMENT{\textit{Rank-1 update} $\mathbf{L}_k\mathbf{L}_k^H = \mathbf{L_{22}}\mathbf{L_{22}}^H+\mathbf{l_{21}}\mathbf{l_{21}}^H$}
			\FOR {$l = 1$ to $N$}
			\STATE $ r \gets \sqrt{[{\mathbf{L_{22}}}]_{l,l}^2 + |[{\mathbf{l}_{21}}]_{l}|^2 }, c \gets \frac{[{\mathbf{L_{22}}}]_{l,l}}{r}$, $ s \gets \frac{[{\mathbf{l}_{21}}]_{l}^*}{r} $
			\STATE $ [{\mathbf{L_{22}}}]_{l:N,l} \gets c[{\mathbf{L_{22}}}]_{l:N,l} + s[{\mathbf{l}_{21}}]_{l:N} $
			\STATE $ [{\mathbf{l}_{21}}]_{l:N} \gets c[{\mathbf{l}_{21}}]_{l:N} - s^*[{\mathbf{L_{22}}}]_{l:N,l} $    	
			\ENDFOR
			\STATE $\mathbf{L}_k \gets \mathbf{L_{22}}$
		\end{algorithmic} 
	\end{algorithm}
	
	\section{Matrix Inversion for Time-Varying Sliding Window Turbo Equalizers}\label{sec:matrixinvcholesk}
	
	\subsection{Shortcomings of Existing Approaches}
	Time-varying FIR as in (\ref{eq_fir_model}) have excessive computational costs due to symbol-wise filter updates, requiring recursive matrix inversion methods. This section overviews the problem of computing $\mathbf{f}_k=\mathbf{\Sigma}^{-1}_k\mathbf{h}_k$, for $k=0,\dots,K-1$ efficiently.
	
	In \cite{tuchler_minimum_2002}, T\"{u}chler et al. propose for LE-IC, a recursive matrix inversion algorithm, based on common submatrices between successive inverses. The procedure requires computing an initial inverse (Gauss-Jordan inversion) with a complexity  order\footnote{"Order" means asymptotic expansion as $N\rightarrow +\infty$, assuming $N\propto 3L$, i.e. sliding window operating on $4L$ symbols.} of $4N^3/3$, but further recursions' complexity is $2N^2$.
	
	Practical implementations avoid inversion by solving the system $\mathbf{\Sigma}_k\mathbf{f}_k=\mathbf{h}_k$ for $\mathbf{f}_k$ with triangular factorizations \cite{studerFateh_ASICimplementationOf}, using forward/backward substitutions. This approach is even more advantageous in equalization where the system is sparse. 
	
	In this paper, we propose a novel recursive inversion strategy for LE-IC and DFE-IC, based on an initial Cholesky decomposition, and followed by sparse rank-1 updates/downdates of the factors for following inversions. 
	Unlike \cite{studerFateh_ASICimplementationOf}, our algorithm is able to deal with channel matrices evolving in time, making it more efficient for turbo TV FIR. For LE-IC the complexity order is of $N^2$, hence roughly 50\% less complex than \cite{tuchler_minimum_2002}.

	\begin{algorithm}[t!]
		\caption{Cholesky update algorithm for DFE-IC.}
		\label{algo:dfecholupdate}
		\begin{algorithmic}[1]
			\renewcommand{\algorithmicrequire}{\textbf{Input}}
			\renewcommand{\algorithmicensure}{\textbf{Output}}
			\REQUIRE $\tilde{\mathbf{L}}_{k},\bar{v}^a_{k-1}, \bar{v}^c_{k-1}, [\mathbf{H}_{k}]_{:,-1}$
			\ENSURE  $\mathbf{L}_{k}$
			\STATE $ \mathbf{w} \gets \sqrt{\vert \bar{v}^a_{k-1}-\bar{v}^c_{k-1}\vert }[\mathbf{H}_{k}]_{:,-1} $
			\FOR {$l = N_p$ to $N$}
			\IF {$\bar{v}^c_{k-1} < \bar{v}^a_{k-1}$}
			\STATE \COMMENT{\textit{Rank-1 downdate} $\mathbf{L}_k\mathbf{L}_k^H = \tilde{\mathbf{L}}_{k}\tilde{\mathbf{L}}_{k}^H-\mathbf{w}\mathbf{w}^H$}
			\STATE $ r \gets \sqrt{[\tilde{\mathbf{L}}_{k}]_{l,l}^2 - |[\mathbf{w}]_{l}|^2 } $, $ c \gets \frac{[\tilde{\mathbf{L}}_{k}]_{l,l}}{r} $, $ s \gets \frac{[\mathbf{w}]_{l}^*}{r} $
			\STATE $ [\tilde{\mathbf{L}}_{k}]_{l:N,l} \gets c[\tilde{\mathbf{L}}_{k}]_{l:N,l} - s[\mathbf{w}]_{l:N} $
			\ELSIF {$\bar{v}^c_{k-1} > \bar{v}^a_{k-1}$}
			\STATE \COMMENT{\textit{Rank-1 update}  $\mathbf{L}_k\mathbf{L}_k^H = \tilde{\mathbf{L}}_{k}\tilde{\mathbf{L}}_{k}^H+\mathbf{w}\mathbf{w}^H$}	
			\STATE $ r \gets \sqrt{[\tilde{\mathbf{L}}_{k}]_{l,l}^2 + |[\mathbf{w}]_{l}|^2 } $, $ c \gets \frac{[\tilde{\mathbf{L}}_{k}]_{l,l}}{r} $, $ s \gets \frac{[\mathbf{w}]_{l}^*}{r} $
			\STATE $ [\tilde{\mathbf{L}}_{k}]_{l:N,l} \gets c[\tilde{\mathbf{L}}_{k}]_{l:N,l} + s[\mathbf{w}]_{l:N} $
			\ENDIF
			\STATE $ [\mathbf{w}]_{l:N} \gets c[\mathbf{w}]_{l:N} - s^*[\tilde{\mathbf{L}}_{k}]_{l:N,l} $    	
			\ENDFOR
			\STATE $\mathbf{L}_k \gets \tilde{\mathbf{L}}_k$
		\end{algorithmic} 
	\end{algorithm}
	
	\subsection{Cholesky Factor Update for MMSE LE-IC}
	
	We consider a LE-IC with priors variances $\bar{v}_k$, let $\mathbf{L}_{k-1}$ be the lower triangular Cholesky decomposition of the covariance matrix $\mathbf{\Sigma}_{k-1}$, i.e. $\mathbf{L}_{k-1}\mathbf{L}_{k-1}^H=\mathbf{\Sigma}_{k-1}$. 
	The resulting updated Cholesky decomposition is a rank-1 update \cite{golub_1996_matrix} of $\mathbf{L}_\mathbf{22}$, defined within algorithm \ref{algo:lecholupdate}.
	
	These steps, followed by forward/backward substitutions $\mathbf{f}_k=\mathbf{L}_k^{-H}\mathbf{L}_k^{-1}\mathbf{h}_k$, allow low complexity filter computation.

	\subsection{Cholesky Factor Update for MMSE DFE-IC}
	
	In the case of DFE-IC, the diagonal of the covariance matrix $\mathbf{\bar{V}}^\text{tdfe}$ is composed of two independently sliding parts: one for causal symbols $\bar{v}^c_k$, between symbols $k-N_p'$ and $k-1$, the other for anti-causal $\bar{v}^a_k$, between symbols $k$ and $k+N_d$. The LE-IC update procedure above handles the addition of $\bar{v}^a_{k+N_d}$ and the removal of $\bar{v}^c_{k-N_p'-1}$, but the change in ${(k-1)}^\text{th}$ symbol remains to be updated.
	
	Algorithm \ref{algo:dfecholupdate} gives a such update procedure for DFE-IC, by applying either a rank-1 update or downdate on $\tilde{\mathbf{L}}_k$, the Cholesky factor who has already been updated by algorithm \ref{algo:lecholupdate}, depending on the sign of $\bar{v}^c_{k-1}-\bar{v}^a_{k-1}$. Such updates are carried out using Givens plane rotations \cite{golub_1996_matrix}.

	\subsection{Computational Complexity Analysis}
	
	The computational complexity of the proposed algorithm is evaluated with the number of required multiply and accumulate units, estimated by the number of real additions and multiplications, amounting to half a floating point operation ($0.5$~FLOPs) each.
	
	\begin{figure}[t!]
		\centering
		\includegraphics[width=3.1in]{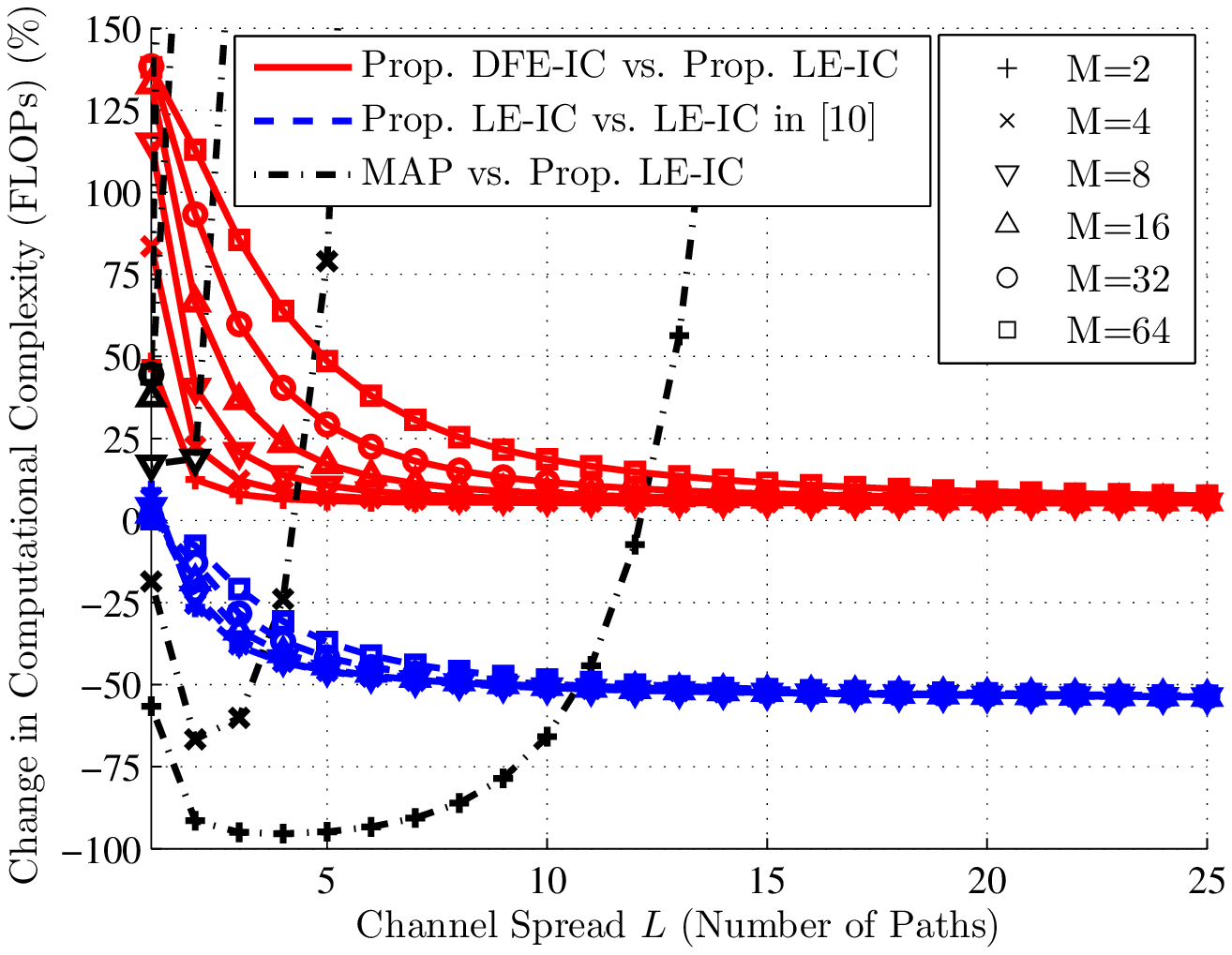}
		\caption{Complexity comparison of LE-IC and DFE-IC with proposed matrix inversion algorithm.}
		\label{theocomplexity_proakisC}
	\end{figure}

	FLOP count ratios between different FIR implementations are plotted in Fig. \ref{theocomplexity_proakisC}, depending on the channel spread, with a block length $K=2048$ and a FIR window given by $N=3L+2$, $N_d=2L$.
	The blue dashed curves show the FLOP count ratio of a LE-IC using our strategy relative to using the algorithm in \cite{tuchler_minimum_2002}, for different constellation orders. Up to 50\% saving is observed as channel spread increases. 
	
	DFE-IC FLOP count is compared to LE-IC, both using the proposed inversion strategies, with red solid lines. This ratio is high for a low number channel taps, but decreases to 7\% as $L$ increases, more or less quickly depending on the modulation order $M$. 
	Finally, MAP detector is seen to be an interesting alternative to FIR receivers for BPSK/QPSK signalling, in channels with very short channel spreads.

	\section{Comparison with the prior work on Time-Varying DFE-IC Structures}\label{sec:dfe}
	
	In this section, the DFE-IC based on EP feedback, proposed in section \ref{ssec:dfeic_ep}, in its canonical form without self-iterations ($\mathcal{S}_\tau=0$) and without damping is compared to alternative state-of-the-art TV DFE-IC structures.

	First, to provide a fair performance comparison with alternatives, existing suboptimal DFE-IC schemes \cite{tuchler_turbo_2002,jeong_turbo_2010,tao_2016_low} are extended to time-varying structures using soft posterior feedback.
	Next, analytical and asymptotic analysis, and Monte Carlo simulations show the superiority of DFE-IC based on EP relative to LE-IC, classical DFE and concurrent DFE-IC structures.

	\subsection{On the TV DFE-IC based on Bayesian estimators}\label{ssec_dfeic_app}
	
	References on time-varying DFE-IC with soft feedback are limited. Hence, here existing methods are generalized and improved before comparison, thanks to our framework, in order to provide a fair comparison.
	Until EP, soft posterior feedback was the only imperfect feedback with a reasonable complexity in the literature, applicable to any constellation. Nevertheless, it is not possible to derive a structure using such feedback within the conventional BP formalism, but here its usage is justified with Bayesian inference. 
	
	One can consider the equalization problem within a Bayesian framework, where a particular realization of a random data symbol is estimated. For instance, the conventional MMSE linear turbo receiver \cite{tuchler_minimum_2002} is also the MAP estimator, if priors are forced to lie in the family of Gaussian distributions \cite{boutros_2002_iterativeMUDCDMA}. Hence this equalizer is the unbiased Bayesian estimator $\mathbb{E}_{\mathcal{L}_k}[x_k \vert \mathbf{y}_k, \mathbf{H}_k]$, where the joint prior distribution
	$\mathcal{L}_k(\mathbf{x}_k) \propto  \prod_{l=k-N_p', }^{k+N_d}\mathcal{CN}({x}^p_l, v^p_l)$
	is used. However, in Bayesian estimation theory, the mean square error can be further reduced, using a \emph{sequential} MMSE estimator, which improves its posterior with previously estimated data (Sect.~12.6 in \cite{mckay_1993_fundermentals_of_stat_signal_proc}). 
	Following this idea, we propose the improved estimator $\mathbb{E}_{\mathcal{A}_k}[x_k \vert \mathbf{y}_k, \mathbf{H}_k]$, based on the joint posterior 
	$\mathcal{A}_k(\mathbf{x}_k) \propto  \prod_{l=k}^{k+N_d}\mathcal{CN}({x}^p_l, v^p_l) \prod_{l=k-N_p'}^{k-1}\mathcal{CN}(\mu^d_l, \gamma^d_l)$,  
	where $\mu^d_l$ and $\gamma^d_l$ are given by (\ref{eq_dem_stats}).
	In the following, we derive a posterior feedback based DFE-IC using this estimator for IC, with model (\ref{eq_fir_model}).

	\subsubsection{Exact TV DFE-IC with APP Feedback} This equalizer is a generalization of invariant schemes in  \cite{lopes_soft-feedback_2006,tao_2016_low} to TV structures.
	It is derived by using the joint posterior $\mathcal{A}_k(\mathbf{x}_k)$ with the model (\ref{eq_fir_model}), derived in the Appendix A.
	The resulting APP FIR structure, is fully defined by
	\begin{equation}
		\begin{split}
			\mathbf{\bar{x}}^\textbf{app}_k &= [\mu^d_{k-N_p'},\dots, \mu^d_{k-1},v^p_k,\dots, x^p_{k+N_d}]^T,\\
			\mathbf{\bar{v}}^\textbf{app}_k &= [\gamma^d_{k-N_p'},\dots, \gamma^d_{k-1},v^p_k,\dots, v^p_{k+N_d}]^T.
		\end{split}
	\end{equation}
	
	This structure will be referred as \emph{DFE-IC APP} in the remainder of this paper, and illustrated in Fig. \ref{fig_dfe_prop} without the dashed module.
	
	\subsubsection{TV DFE-IC with Perfect APP Feedback} Here we propose to generalize \cite{tuchler_turbo_2002,balakrishnan_mitigation_1999} to APP feedback, with perfect decision hypothesis. This imposes decision covariances to 0, focusing the MMSE filter design to only mitigate anti-causal symbol interference. However, its use of hard feedback, i.e. $\arg\max_\alpha \mathcal{D}_k(\alpha)$, was shown to be seriously prone to error propagation \cite{tuchler_turbo_2002}. While \cite{balakrishnan_mitigation_1999} showed improvements with soft posterior feedback on non-turbo, invariant structures, here, we extend this case to time-varying turbo structures. 
	
	This case named \emph{DFE-IC PAPP}, differs from the DFE-IC APP with the variance estimates:
	\begin{equation}
		\begin{split}
			\mathbf{\bar{x}}^\textbf{papp}_k &= \mathbf{\bar{x}}^\textbf{app}_k ,\\
			\mathbf{\bar{v}}^\textbf{papp}_k &= [\mathbf{0}_{N_p'}^T,v^p_k,\dots, v^p_{k+N_d}]^T.
		\end{split}
	\end{equation}

	\subsubsection{Hybrid TV DFE-IC with APP Feedback} This structure is an extension of the TV structure from \cite{jeong_turbo_2010} to APP feedback. In \cite{jeong_turbo_2010}, the DFE-IC with perfect hard decisions from \cite{tuchler_turbo_2002} is improved by adding an estimate of the decision error to the equalizer output variance $v^e_k$. This quantity is given by $\text{Var}_{\mathcal{D}_k}[\mathbf{g}_k^c{}^H([\mathbf{x}-\pmb{\mu}^\mathbf{d}]_{k-N_p':k-1})]$, using (\ref{eq_posteriors}). Moreover, this structure checks whether this variance causes sign changes in extrinsic LLRs, and sets ambiguous LLRs to zero.
	
	This receiver is extended to use APP soft feedback, instead of hard decisions, and denoted \emph{DFE-IC HAPP}.

	\begin{figure}[t]
		\centering
		\includegraphics[width=3.1in]{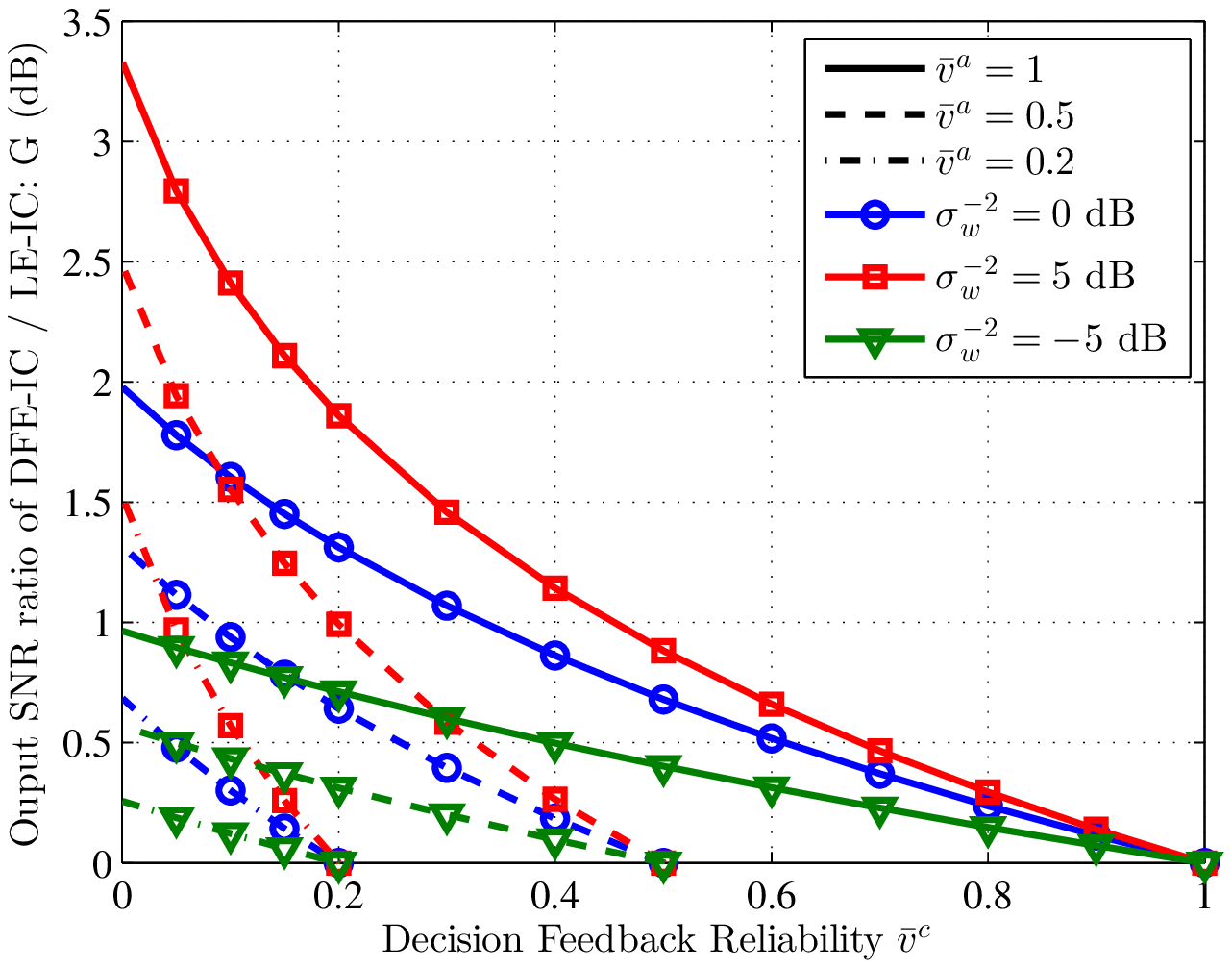}
		\caption{Post-equalization SNR ratio $G$ depending on channel SNR $\sigma_w^{-2}$, prior reliability $\bar{v}^a$ and  ``decision" reliability $\bar{v}^c$.}
		\label{proakisC_theoSNR}
	\end{figure}

	\subsection{Analytic Comparison of DFE-IC vs. LE-IC}
	\label{ss_rapidanalysis}
	
	This paragraph semi-analytically assesses the behaviour of a DFE-IC relative to a LE-IC to underline the interest in jointly using decision feedback and prior information for IC.
	
	In fact, LE-IC operating with priors $(\bar{x}_k,\bar{v}_k)$ provides a lower bound for the achievable information rate of a DFE-IC structure using the same prior information for its anti-causal symbols $(\bar{x}^\text{a}_k,\bar{v}^\text{a}_k)=(\bar{x}_k,\bar{v}_k)$, alongside decision feedback estimates $(\bar{x}^\text{c}_k,\bar{v}^\text{c}_k)$ (see (\ref{eq_dfeic})). 
	By exploiting the structural similarities between DFE-IC and LE-IC, the causal feedback's impact is reflected on a ratio of post-equalization SNR\footnote{$\text{SNR}_\text{out}^\text{XX}=\sigma_x^2/\mathbb{E}[v_k^{e(\text{XX})}]$ is the post-equalization SNR,  where $\text{XX}$ is ``$\text{dfe}$" or ``$\text{le}$",  (see (\ref{eq_fir_model}) for $v^e_k$). Superscript ``le" refers to the use of $(\bar{x}_k,\bar{v}_k)$ for IC, and ``dfe" refers to the use of $(\bar{x}^\text{a}_k,\bar{v}^\text{a}_k)$ and $(\bar{x}^\text{a}_c,\bar{v}^\text{a}_c)$ for IC, as in (\ref{eq_dfeic}).}
	\begin{equation}
		G = \frac{\text{SNR}_\text{out}^\text{dfe}}{\text{SNR}_\text{out}^\text{le}} = \frac{\sigma_x^2}{\mathbb{E}[v_k^{e(\text{dfe})}]}\frac{\mathbb{E}[v_k^{e(\text{le})}]}{\sigma_x^2} = \frac{\xi^\text{dfe}}{\xi^\text{le}}\frac{1-\bar{v} \xi^\text{le}}{1-\bar{v}\xi^\text{dfe}} \label{eq_snrgain}
	\end{equation}
	where $\bar{v}=\mathbb{E}[\bar{v}_k]$ and $\xi^\text{XX}=\mathbb{E}[\xi^\text{XX}_k]$, where $\text{XX}$ is $\text{``le"}$ or $\text{``dfe"}$. This gain is greater than unity iff $\xi^\text{dfe} \geq \xi^\text{le}$, or equivalently iff $\mathbb{E}[\mathbf{\bar{V}}^\textbf{le}_k-\mathbf{\bar{V}}^\textbf{dfe}_k]$ is positive semi-definite. Hence having $\bar{v}>\bar{v}^c$, $\bar{v}^c=\mathbb{E}[\bar{v}^c_k]$ for DFE-IC is required for achieving improvements. 
	Based on empirical and experimental evidence not presented here, the conjecture $\mathbb{P}[\bar{v}_k^c>\bar{v}_k]<0.5$ has been verified over a wide range of input SNRs, and for random constellations, for $\bar{v}^c_k=v^d_k$ (DFE-IC EP) and for $\bar{v}^c_k=\gamma^d_k$ (DFE-IC APP). This ensures $\bar{v}>\bar{v}^c$ and thus, LE-IC output SNR is a lower bound on DFE-IC EP/APP, as possible detection degradations are small. 
	
	$G$ is plotted in Fig. \ref{proakisC_theoSNR}, with $N=17, N_d=10$ and $\sigma_x^2=1$ for the static Proakis-C channel, $\mathbf{h}=[1,2,3,2,1]/\sqrt{19}$; when decisions are more reliable than priors, $G$  increases, otherwise DFE-IC brings small improvements. When $\bar{v}^a \rightarrow 1$, there is no prior information, and decisions bring a significant gain. Oppositely, when $\bar{v}^a \rightarrow 0$, prior information is already close to the ideal, and DFE-IC cannot improve further.  This indicates boosted performance at initial turbo-iterations.

	\ifdouble
	\begin{figure}[t!]
		\centering
		\includegraphics[width=3.2in]{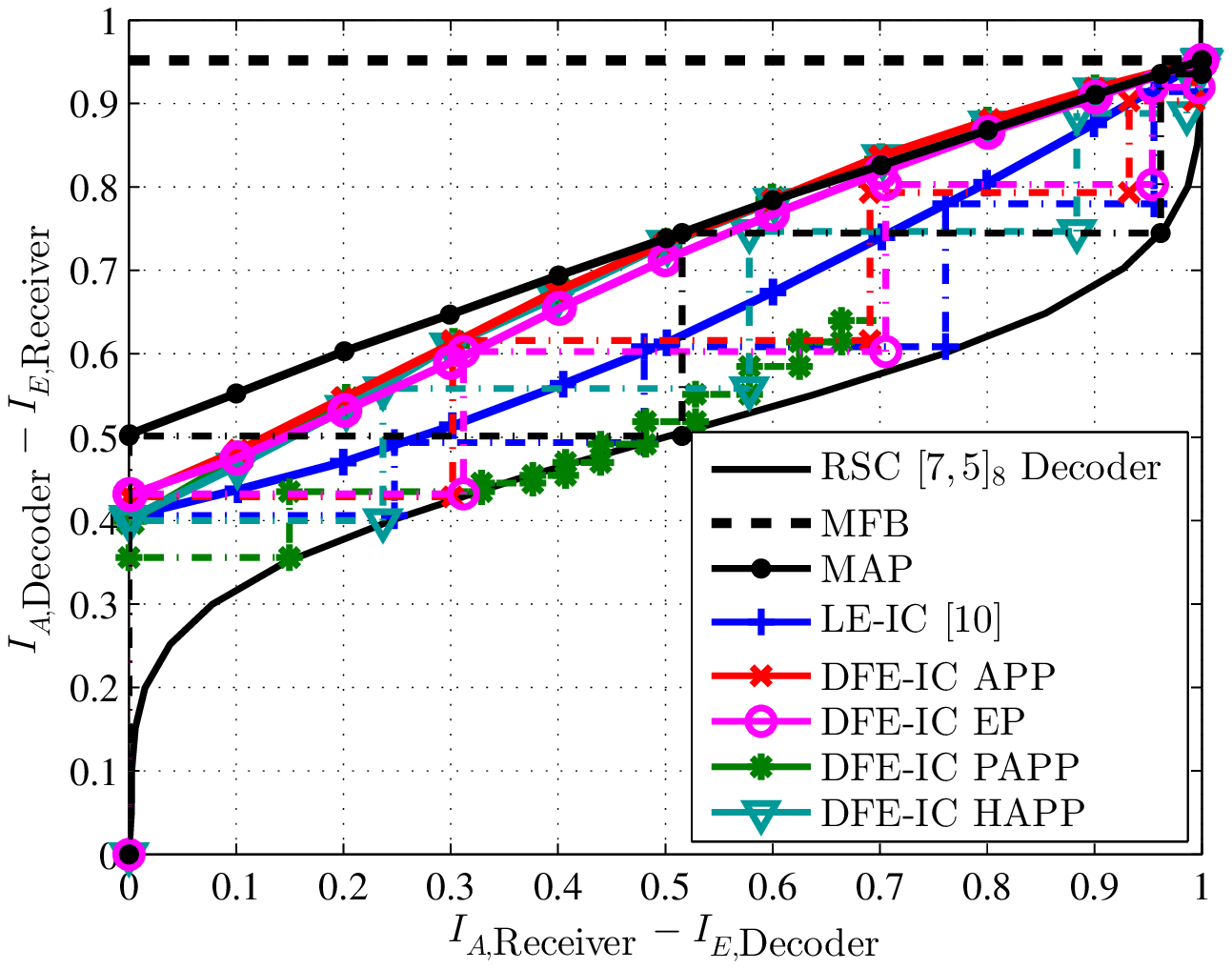}
		\caption{EXIT curves and average MI trajectories of FIR equalizers with BPSK in Proakis C channel at $E_b/N_0=7$dB.}
		\label{bpskProakisC_EXIT_DFEIC}
	\end{figure}
	\else
	\begin{figure}[t!]
		\centering
		\includegraphics[width=3in]{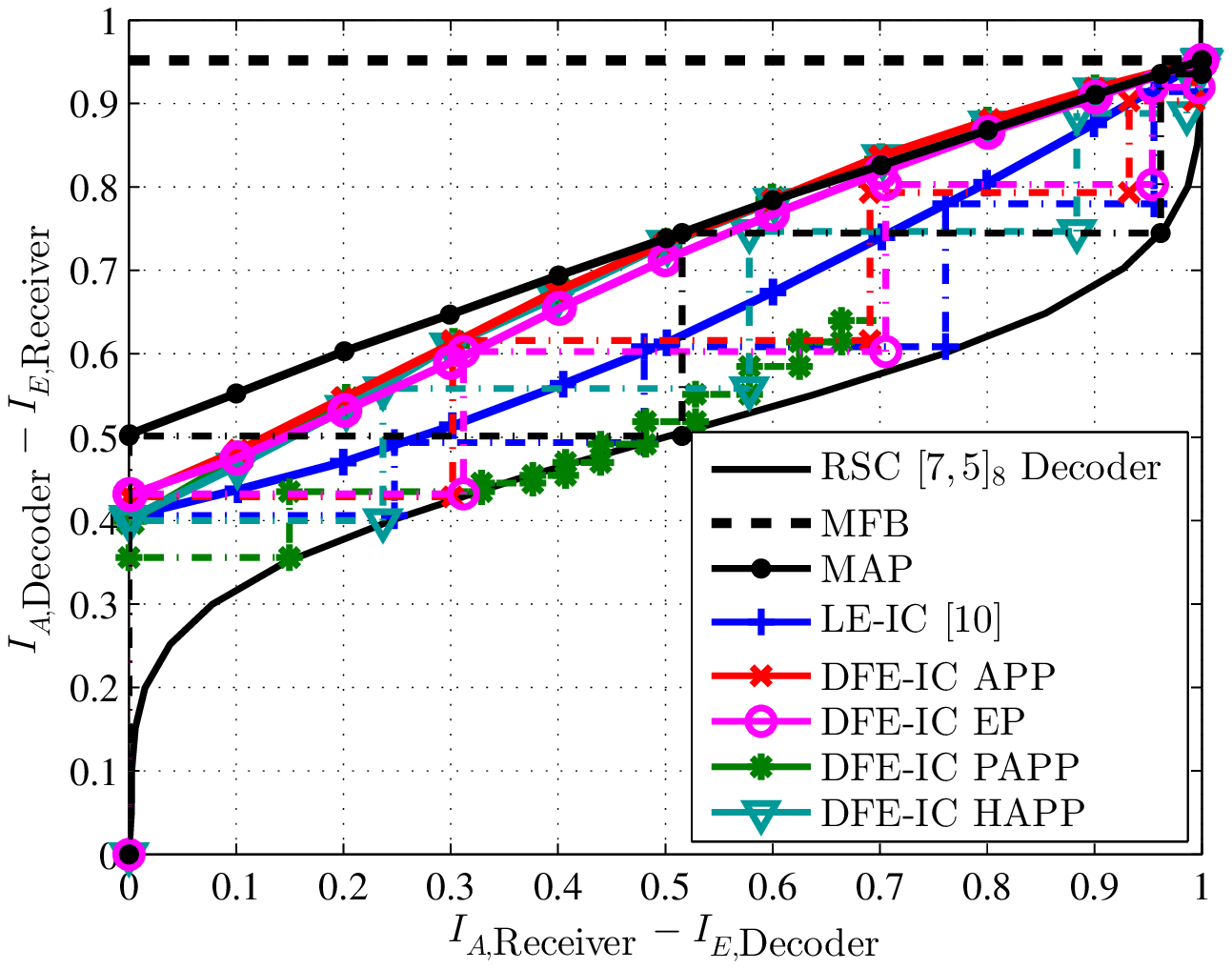}
		\caption{EXIT curves and average MI trajectories of FIR equalizers with BPSK in Proakis C channel at $E_b/N_0=7$dB.}
		\label{bpskProakisC_EXIT_DFEIC}
	\end{figure}
	\fi

	\ifdouble
	\begin{figure}[t!]
		\centering
		\includegraphics[width=3.2in]{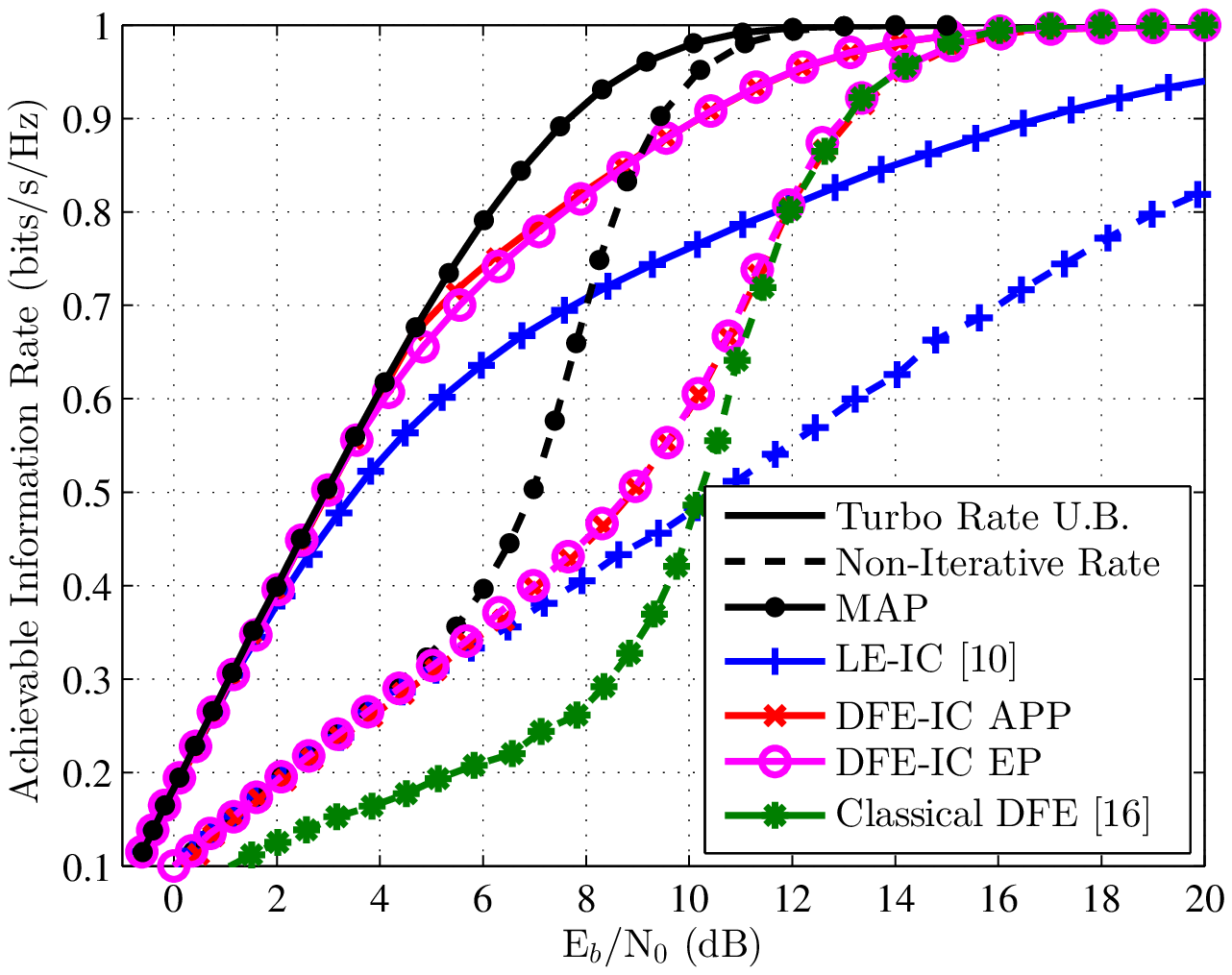}
		\caption{Achievable spectral efficiency on deterministic Proakis C channel with BPSK.}
		\label{fig_effSpecProakisCIntro}
	\end{figure}
	\else
	\begin{figure}[t!]
		\centering
		\includegraphics[width=3in]{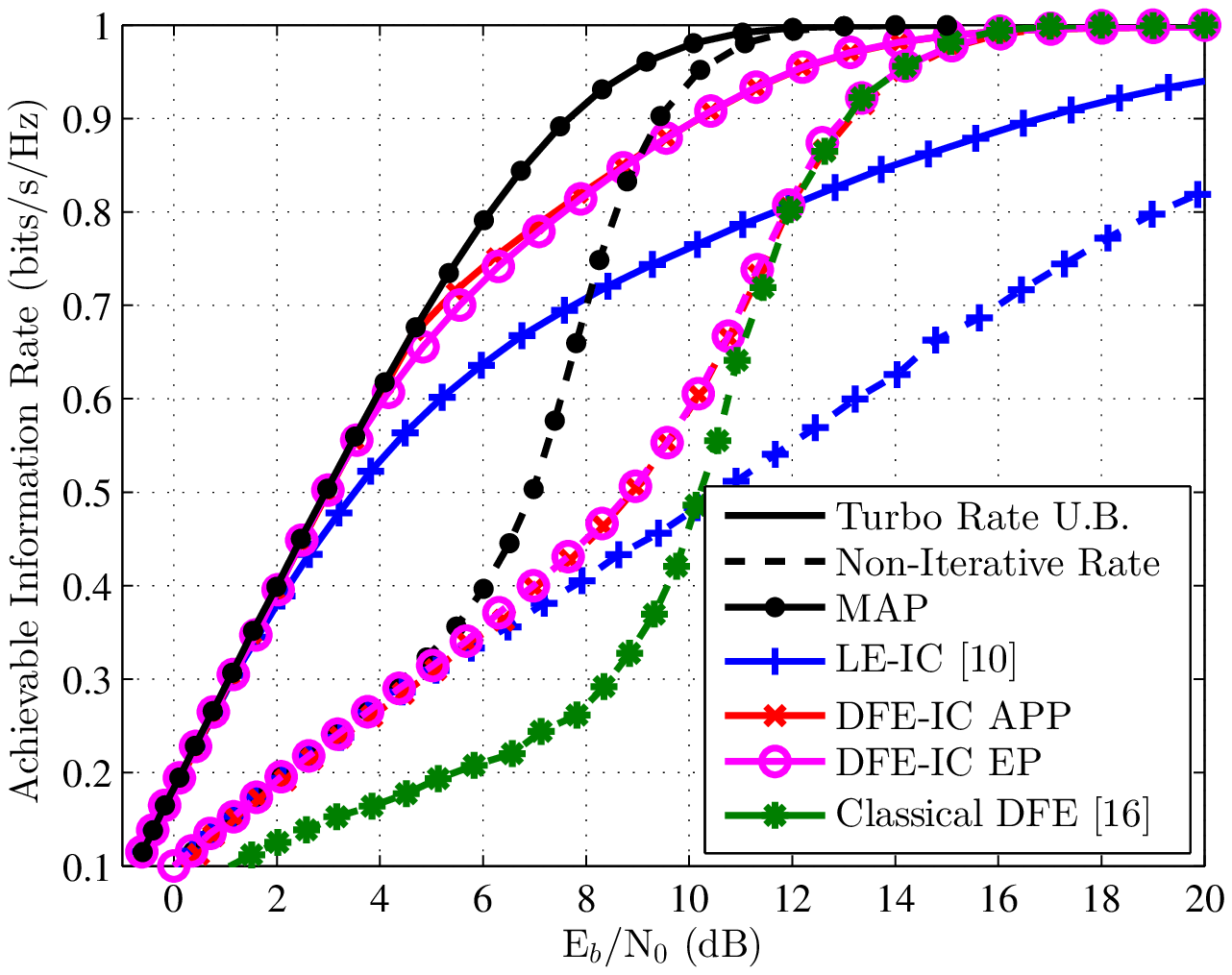}
		\caption{Achievable spectral efficiency on deterministic Proakis C channel with BPSK.}
		\label{fig_effSpecProakisCIntro}
	\end{figure}
	\fi

	\ifdouble
	\begin{figure*}
		\centering
		\includegraphics[width=7in]{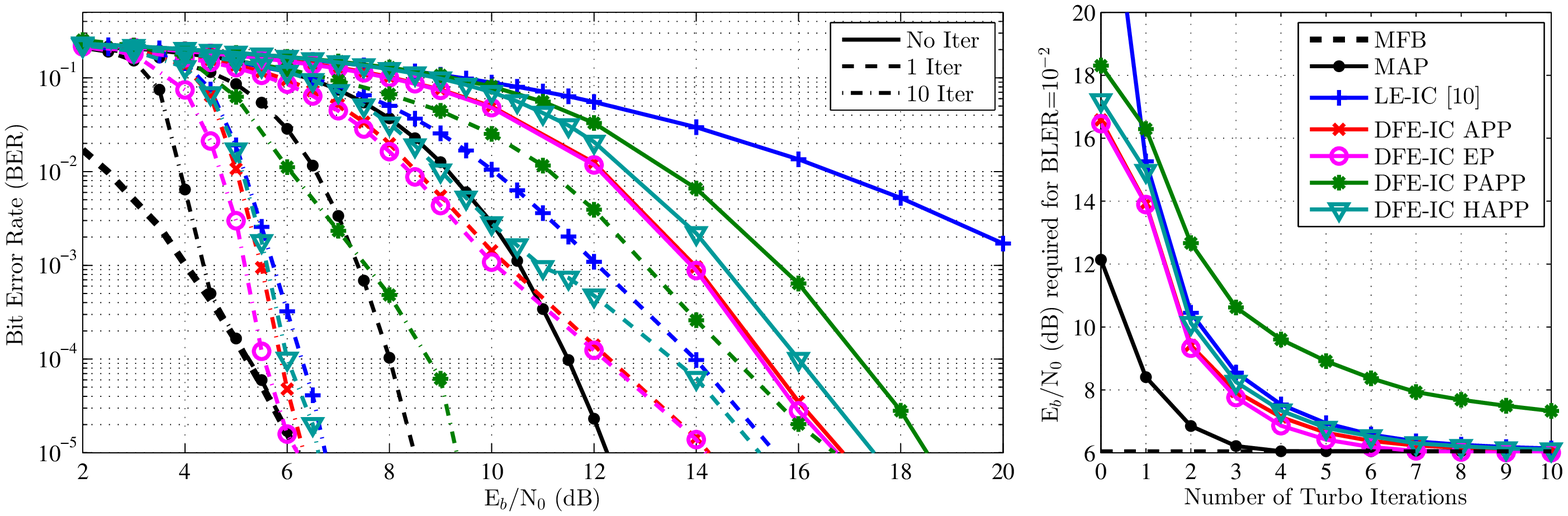}
		\caption{BER and convergence performance of the proposed DFE-IC in Proakis C channel with BPSK constellation.}
		\label{berandConvergenceBPSK_proakisC}
	\end{figure*}
	\else
	\begin{figure*}
		\centering
		\includegraphics[width=6.5in]{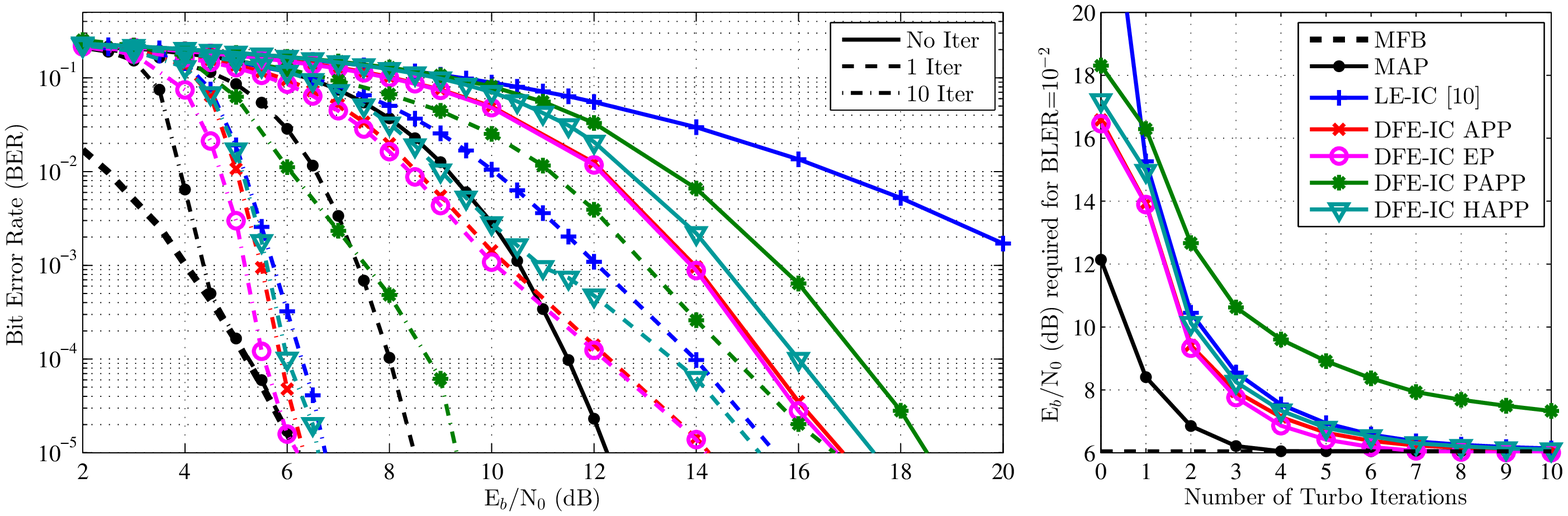}
		\caption{BER and convergence performance of the proposed DFE-IC in Proakis C channel with BPSK constellation.}
		\label{berandConvergenceBPSK_proakisC}
	\end{figure*}
	\fi

	\ifdouble
	\begin{figure*}[t!]
		\centering
		\includegraphics[width=7in]{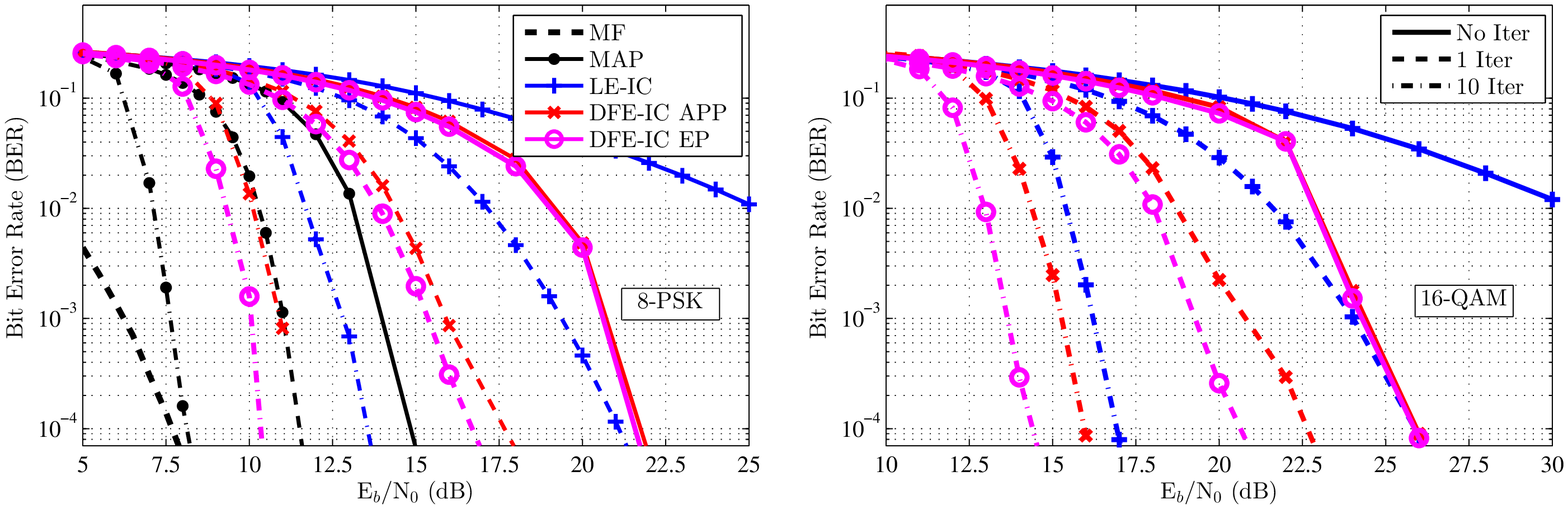}
		\caption{BER performance of the proposed DFE-IC in Proakis-C with 8-PSK and 16-QAM constellations.}
		\label{ber8PSK16QAM_proakisC}
	\end{figure*}
	\else
	\begin{figure*}[t!]
		\centering
		\includegraphics[width=6.5in]{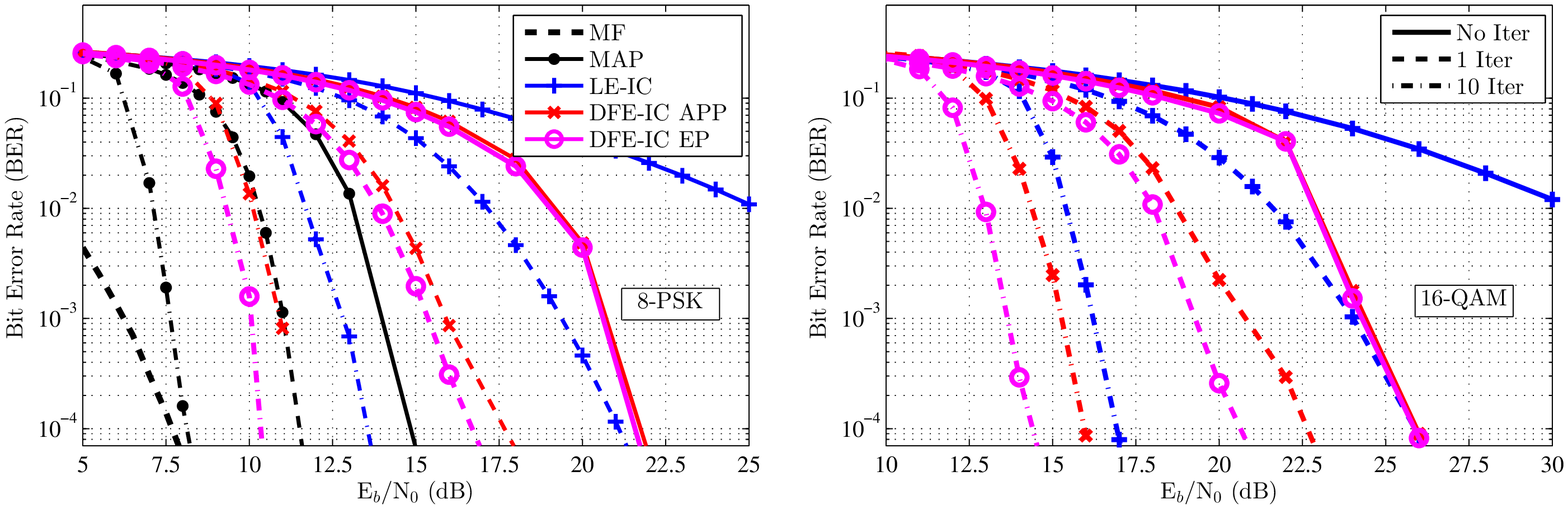}
		\caption{BER performance of the proposed DFE-IC in RSC-coded Proakis-C channel.}
		\label{ber8PSK16QAM_proakisC}
	\end{figure*}
	\fi

	\subsection{Asymptotic Analysis and Performance Prediction}\label{subsec_asymanalysis}

	To assess the full potential of DFE-IC, asymptotic analysis is used to evaluate its achievable rates. Extrinsic information transfer (EXIT) analysis \cite{brink_2000_designing} of a SISO module is used as a tool for characterizing its asymptotic limits, by tracking extrinsic mutual information (MI) exchanges between the iterative components. 
	Essentially, a SISO receiver can be characterized by a simple transfer function $I_E=\mathcal{T}_R(I_A,\mathbf{H},\sigma_w^2)$, where $I_A$ and $I_E$ are the MI between coded bits and respectively its input prior LLRs and output extrinsic LLRs, and $\sigma_w^2$ and $\mathbf{H}$ show its dependence on the channel and the received SNR.

	In Fig. \ref{bpskProakisC_EXIT_DFEIC}, transfer curves $\mathcal{T}_R$ are plotted in solid lines for considered receivers along with the reverse transfer $\mathcal{T}_D^{-1}$ of the BCJR decoder of a recursive systematic convolutional (RSC) code. DFE-IC APP yields a higher $I_R$ than LE-IC for all $I_A$, unsurprisingly given the posterior feedback, and there is little difference with DFE-IC EP, which has slightly lower rates at low prior information.
	In particular, the improvement at $I_A=0$ lets us conjecture a lower waterfall threshold in BPSK, and the higher slope of the $\mathcal{T}_R$ curve at low $I_A$ hints an improved convergence speed across turbo iterations. 
	EXIT curves provide a fairly accurate waterfall threshold estimation and can be used for code design \cite{narayanan_2005_estimatingPDF}.
	
	Another use of EXIT analysis is performance prediction, however this involves strong assumptions on prior inputs that often cannot be met for FIR turbo equalizers in practice. 
	Hence, EXIT curves only provide an upper-bound on information rate for receivers other than MAP. In this respect, it is then interesting to compare transfer curves, with actual MI trajectories (in dashed lines in Fig. \ref{bpskProakisC_EXIT_DFEIC}).

	It had been noted in \cite{tuchler_turbo_2002}, that trajectories of DFE-IC with hard, ``perfect" decision assumption do not follow EXIT curves; this issue remains with DFE-IC PAPP, although less severely, indicating that the ``perfect decisions" assumption causes a severe information loss.
	Other FIRs' trajectories overall follow receiver and decoder curves and reach MFB, but after a few iterations, they no longer make contact with transfer curves, losing convergence speed. This is a common disadvantage of FIR equalizers, attributed to  short cycles caused by neighbouring symbol correlations, 
	as shown in Fig. 16 in \cite{tuchler_turbo_2002}. 
	However note that among DFE-IC receiver, EP feedback yields trajectories that remains closest to EXIT curves, making it easier to predict.

	The achievable spectral efficiency for a given receiver can be measured with the help of the area theorem for EXIT charts \cite{hagenauer_exit_2004}.
	In Fig. \ref{fig_effSpecProakisCIntro}, achievable rates for BPSK constellation are plotted. Note that for MAP receivers, this rate is an accurate approximation of the channel symmetric information rate (SIR)
	\cite{arnold_2006_simulation}. As non-iterative FIR do not depend on prior inputs, their achievable rates are also accurately computed. For turbo FIR, upper bounds are obtained by combining results of area theorem with the channel SIR. Tightness of this bound depend on the closeness of true MI trajectories to EXIT charts in Fig. \ref{bpskProakisC_EXIT_DFEIC}, so APP feedback's asymptotic performance is likely to be overestimated compared to EP feedback.
	
	\subsection{Finite-Length Comparison with Existing Schemes}
	\label{ssec_dfeicberanalysis}
	
	Monte Carlo integration remains the most reliable analysis approach joint detection of BPSK symbols is considered with parameters in section \ref{ss_rapidanalysis}, and $K_u=2048$, coded with a terminated $[7,5]_8$ RSC code. Bit error rate (BER) of various receivers are plotted in Fig. \ref{berandConvergenceBPSK_proakisC}. For the reported iterations, the DFE-IC APP outperforms other APP feedback DFE structures,
	and their convergence speeds are compared on the right side of the figure, at a block error rate (BLER) of $10^{-2}$. 
	EP-based feedback provides further improvement of the threshold by 0.5 dB relative to APP, and it is shown to reach MFB limit within 7 iterations, earlier than DFE-IC APP.
	
	Assessing DFE-IC performance at low spectral efficiency conditions, as above, is of interest, to remedy the poor behaviour of classical DFE at those operating points (see Fig. \ref{fig_effSpecProakisCIntro}). Indeed, turbo processing helps DFE structures to outperform LE at all rates. A higher spectral efficiency case is plotted on the left side of the Fig. \ref{ber8PSK16QAM_proakisC}, with $8$-PSK constellation in the same configuration; DFE-IC APP is shown to improve LE-IC waterfall threshold by $2$dB, DFE-IC EP asymptotically provides an additional $1.2$dB. On the right side of the Fig. \ref{ber8PSK16QAM_proakisC}, 16-QAM is considered; showing that DFE-IC EP provides further performance enhancements for one or more iterations.
	
	Finally, the coded performance of DFE-IC is balanced with complexity considerations. In Fig. \ref{perfComplexityTradeoff_proakisC}, the receiver computational complexity (FLOPs per symbol) required to decode with a BLER of $10^{-2}$ is plotted as a function of $E_b/N_0$. These values are computed, assuming the use of the proposed matrix inversion algorithm in section \ref{sec:matrixinvcholesk}, and by accounting for the equalization, the demapping and the decoding costs. A curve represents the evolution of BLER and the computational costs of a receiver accross turbo iterations.
	
	DFE-IC provides a better trade-off than LE-IC; at any given complexity, it is more efficient, especially at initial iterations, and the asymptotic $E_b/N_0$ gap between LE-IC and DFE-IC increases with the modulation order $M$. The use of EP feedback is more advantageous at higher iterations, for higher order constellations, while APP is more efficient for non-iterative receivers.

	\ifdouble
	\begin{figure}[t!]
		\centering
		\includegraphics[width=3.2in]{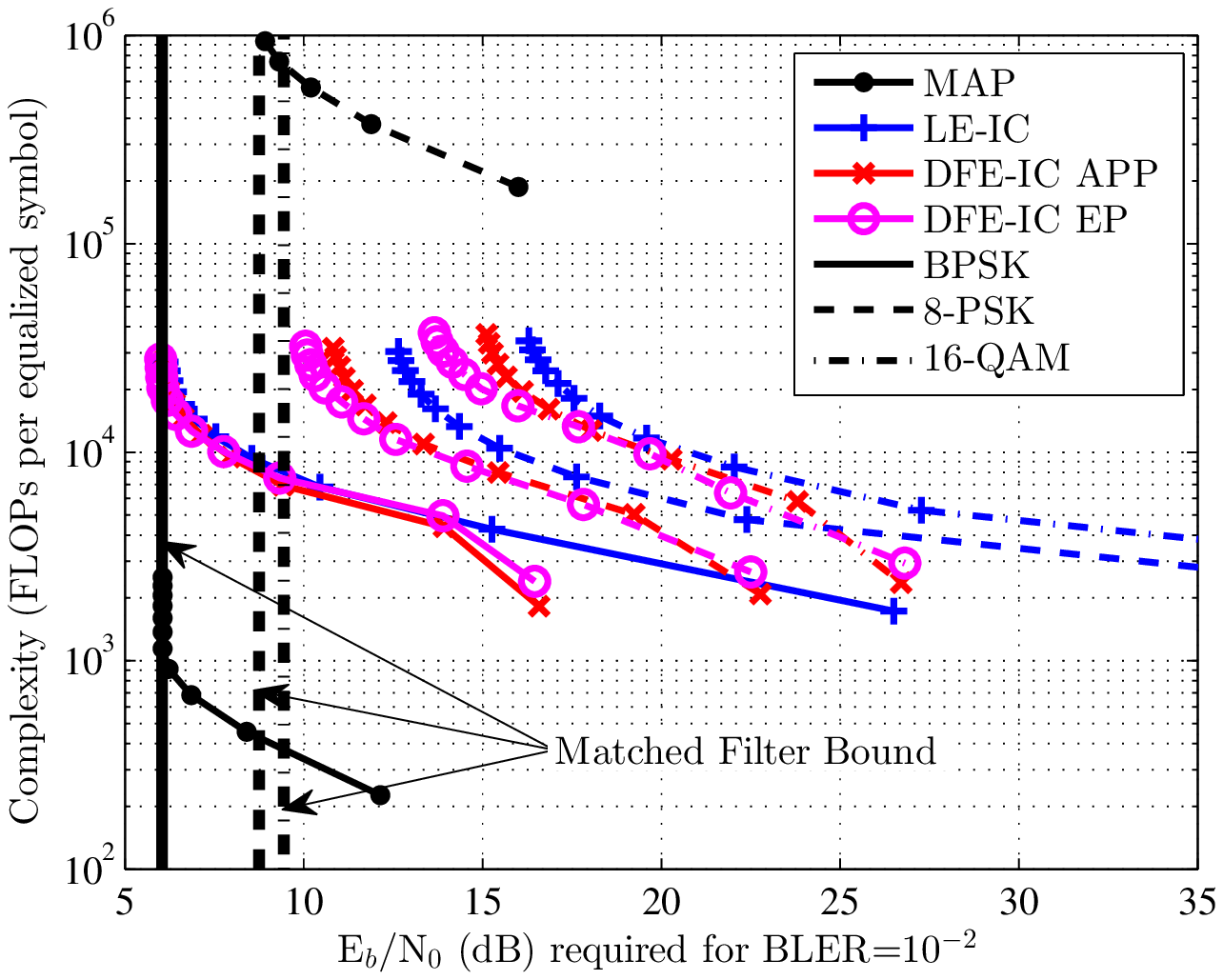}
		\caption{Performance complexity trade-off in Proakis C.}
		\label{perfComplexityTradeoff_proakisC}
	\end{figure}
	
	\else
	\begin{figure}[t!]
		\centering
		\includegraphics[width=3.15in]{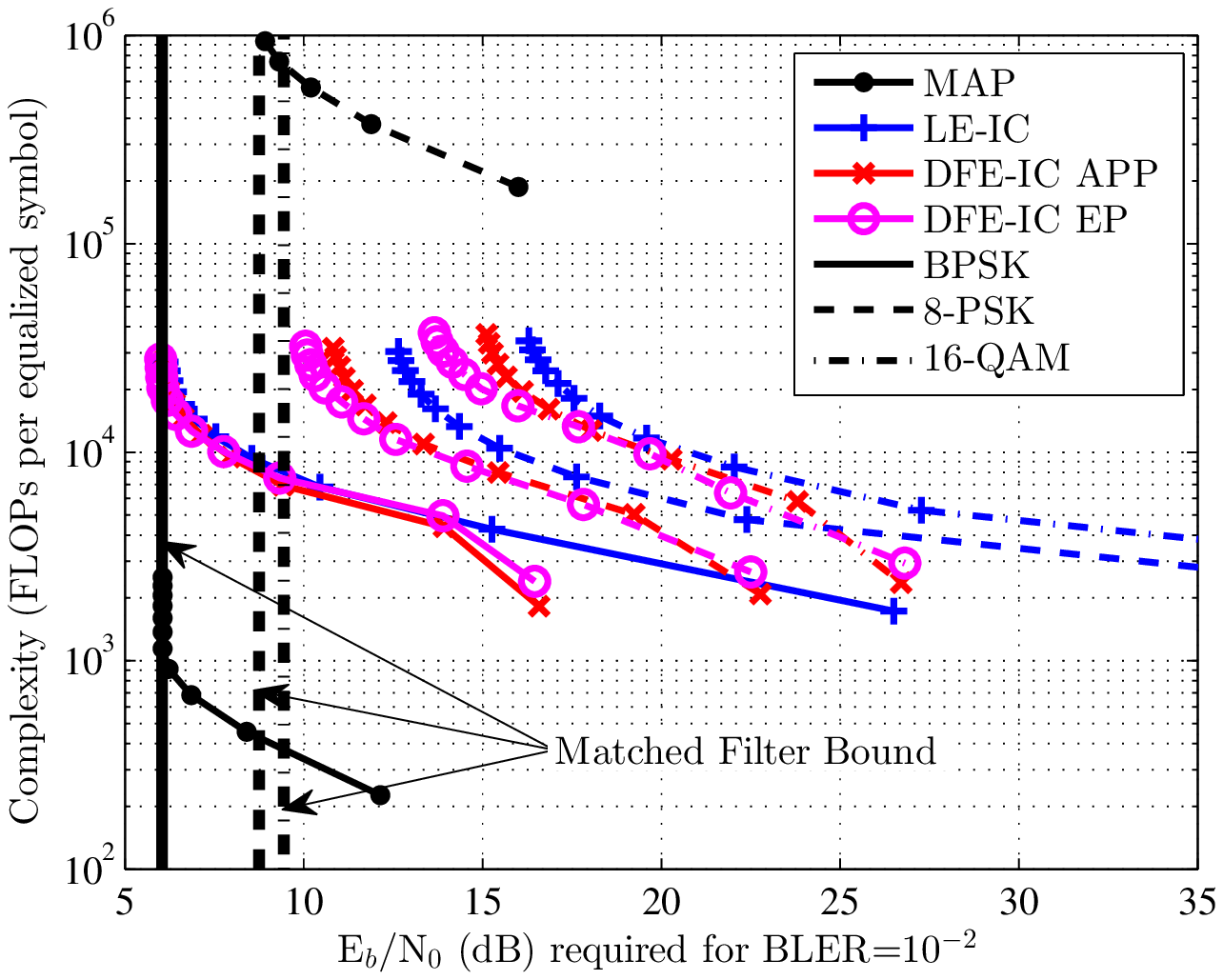}
		\caption{Performance complexity trade-off in Proakis C.}
		\label{perfComplexityTradeoff_proakisC}
	\end{figure}
	\fi

	In conclusion, DFE-IC outperforms LE-IC in various aspects: it converges faster towards MFB, has a lower decoding threshold than LE-IC, especially at higher spectral efficiencies. 
	Among DFE-IC with APP feedback, exact derivation DFE-IC APP is superior according to both finite-length and asymptotic analysis. 
	Although EXIT charts show little difference between DFE-IC EP and APP, in practical simulations EP feedback tends to outperform APP. This is justified by the tightness of EP MI trajectories to EXIT curves; APP is overestimated.
	Although it DFE-IC EP appears to be able to reach channel SIR at low to medium spectral efficiencies, there is still a gap to MAP performance. 
	
	In the following, the use of self-iterations will be assessed to further improve performances.

	\section{Comparison with the prior work on Self-Iterated EP Structures}\label{sec:self}
	
	Some recent EP-based receivers \cite{senstAscheid_2011_frameworkEP_MMSEMIMO, santosMurilloFuentes_2017_EPBLE, santosMurillosFuentes_2017_smoorthingEP,santosMurilloFuentes_2017_EPnuBLE,sahinCipriano_2018_FDSILEEP} have observed remarkable performance improvements in repeating the detection process in a parallel schedule through self-iterations. As the demapping process is computationally less intensive than channel decoding, such structures are of practical interest. In this section, the benefits in using a self-iterated DFE-IC EP compared to structures in prior work is investigated.

	Independently of our work, an EP-based FIR structure is derived in the concomitant work \cite{santosMurilloFuentes_2017_EPnuBLE}. Unlike the message passing formalism used in section \ref{sec:ep_modelling}, structure in \cite{santosMurilloFuentes_2017_EPnuBLE} is obtained by approximating a self-iterated block receiver, derived by EP-based approximation of the posterior PDF (\ref{eq_postb}).  The resulting FIR structure uses a \emph{parallel} schedule, and corresponds to a LE-IC within each SI. Using our formalism, it is equivalent to updating, all VNs $x_k$ with messages from EQU sequentially, and only then activating DEM to update posterior approximations. This process is then iterated with DEM sending back an extrinsic message to EQU, and finally DEM computes messages towards DEC. In the following, the structure denoted as ``EP-F" in \cite{santosMurilloFuentes_2017_EPnuBLE}, is refered as a self-iterated LE-IC (SI LE-IC), with following mean and variances used for IC
	\ifdouble
	\begin{equation}
		\begin{split}
			\mathbf{\bar{x}}^\textbf{le-ep}_k{}^{(s)} &= [{x}^{d(s)}_{k-N_p'},\dots, {x}^{d(s)}_{k+N_d}]^T,\\
			\mathbf{\bar{v}}^\textbf{le-ep}_k{}^{(s)} &= [v^{d(s)}_{k-N_p'},\dots, v^{d(s)}_{k+N_d}]^T.
		\end{split}
	\end{equation}
	\else
	\begin{equation}
		\mathbf{\bar{x}}^\textbf{le-ep}_k{}^{(s)} = [{x}^{d(s)}_{k-N_p'},\dots, {x}^{d(s)}_{k+N_d}]^T,\,
		\mathbf{\bar{v}}^\textbf{le-ep}_k{}^{(s)} = [v^{d(s)}_{k-N_p'},\dots, v^{d(s)}_{k+N_d}]^T.
	\end{equation}
	\fi
	If the computations of messages on EQU is carried out only once ($\mathcal{S}_\tau=0$), this receiver yields the same result as the conventional turbo LE-IC \cite{tuchler_minimum_2002}.

	\ifdouble
	\begin{figure*}[t!]
		\centering
		\includegraphics[width=7in]{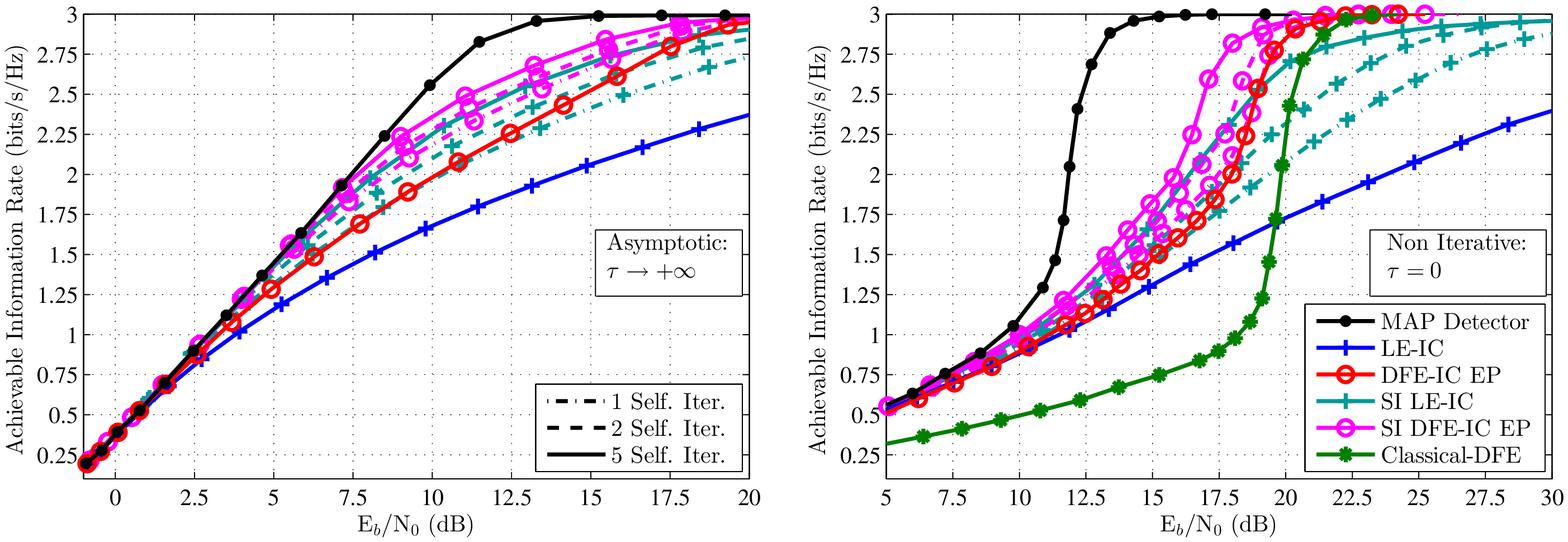}
		\caption{Achievable Rates of Self-iterated LE-IC and DFE-IC in Proakis-C with 8-PSK constellation.}
		\label{fig_symp_8PSK_SIFIR}
	\end{figure*}
	\else
	\begin{figure*}[t!]
		\centering
		\includegraphics[width=6.5in]{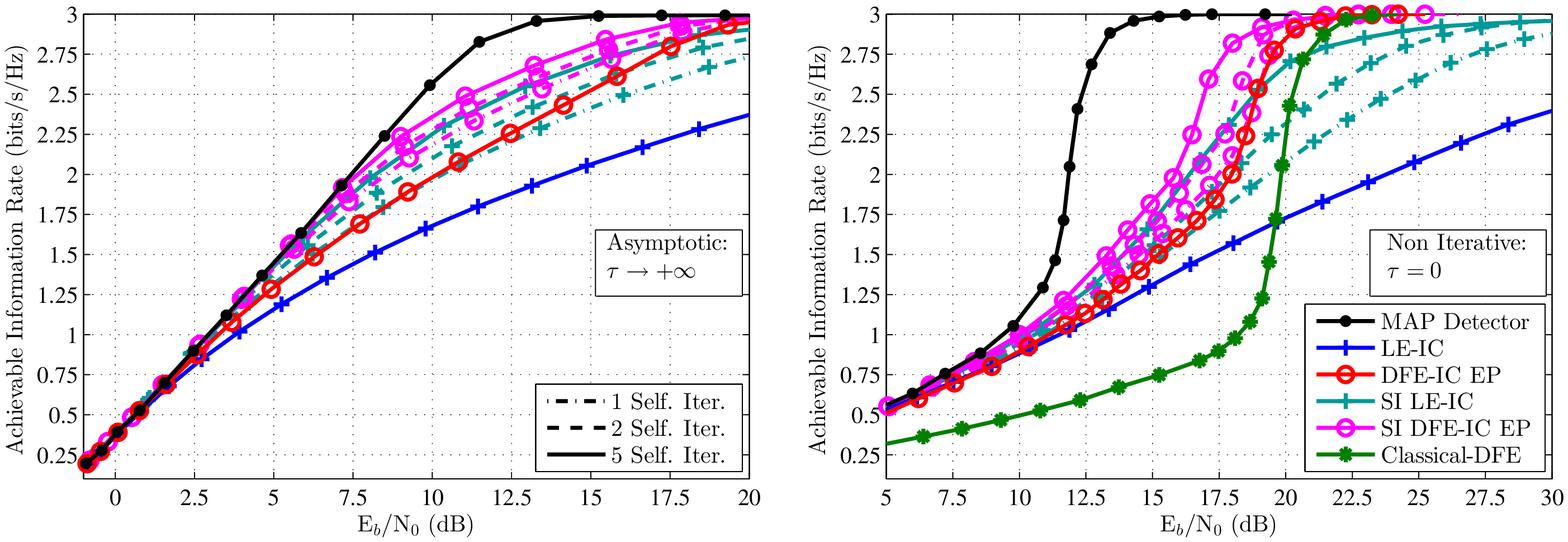}
		\caption{Achievable rates of self-iterated FIR in Proakis-C with 8-PSK constellation.}
		\label{fig_symp_8PSK_SIFIR}
	\end{figure*}
	\fi

	\subsection{Asymptotic Comparison}
	
	First, we look into the achievable rates of SI LE-IC and DFE-IC EP to identify operating points where self-iterations have an advantage.
	
	We consider 8-PSK signalling on the Proakis-C channel, and use the area theorem to obtain an upper bound on asymptotic achievable rates (i.e. $\tau \rightarrow \infty$), plotted on the left side of Fig. \ref{fig_symp_8PSK_SIFIR}. Information rates of the optimal MAP detector, LE-IC and DFE-IC EP without SI, and SI LE-IC and SI DFE-IC are considered. For self-iterated receivers, a static damping with $\beta=0.6$ is used.
	Numerical results show that SI is not required for LE-IC up to 0.75~bits/s/Hz (i.e. using a code rate less than 1/4), as LE-IC is close to the SIR, whereas DFE-IC EP continues to follow MAP rates up to 1~bit/s/Hz (up to a code rate of 1/3). On the other hand, when using 5 self-iterations,  DFE-IC EP follows MAP rates within $0.5$~dB up to 2.25~bits/s/Hz, while LE-IC follows up to 1.85~bits/s/Hz. It is also interesting to note that DFE-IC EP with 2 SI outperforms LE-IC with 5 SI, at all rates, indicating at faster convergence of DFE-IC EP towards asymptotic limits.
	
	At the right side of Fig.  \ref{fig_symp_8PSK_SIFIR}, non-turbo iterative achievable rates of these receivers, and those of the classical DFE \cite{belfiore_79_DFE}, are compared. These rates are accurate, and not an upper bound, unlike asymptotic rates, and note that MAP detector is a mere maximum likelihood (ML) detector in this case.  Although self-iterations significantly improve LE-IC performance, at rates above 2.75 bits/s/Hz, classical DFE still outperforms these receivers. DFE-IC EP on the other hand outperforms alternative FIRs at any given self iteration. 
	
	Note that the gap to capacity still remains significant for non turbo iterative rates, and to some extent, for asymptotic rates. Hence with the objective of deriving capacity achieving practical receivers in mind, future work should explore the usage of the proposed DFE-IC EP as a constituent element for bidirectional DFE \cite{jeong_soft-soft-out_2011} or for concatenated FIR  \cite{jeongMoon_2013_selfiteratingSoftEqualizer} receivers.

	\ifdouble
	\begin{figure*}[t!]
		\centering
		\includegraphics[width=7in]{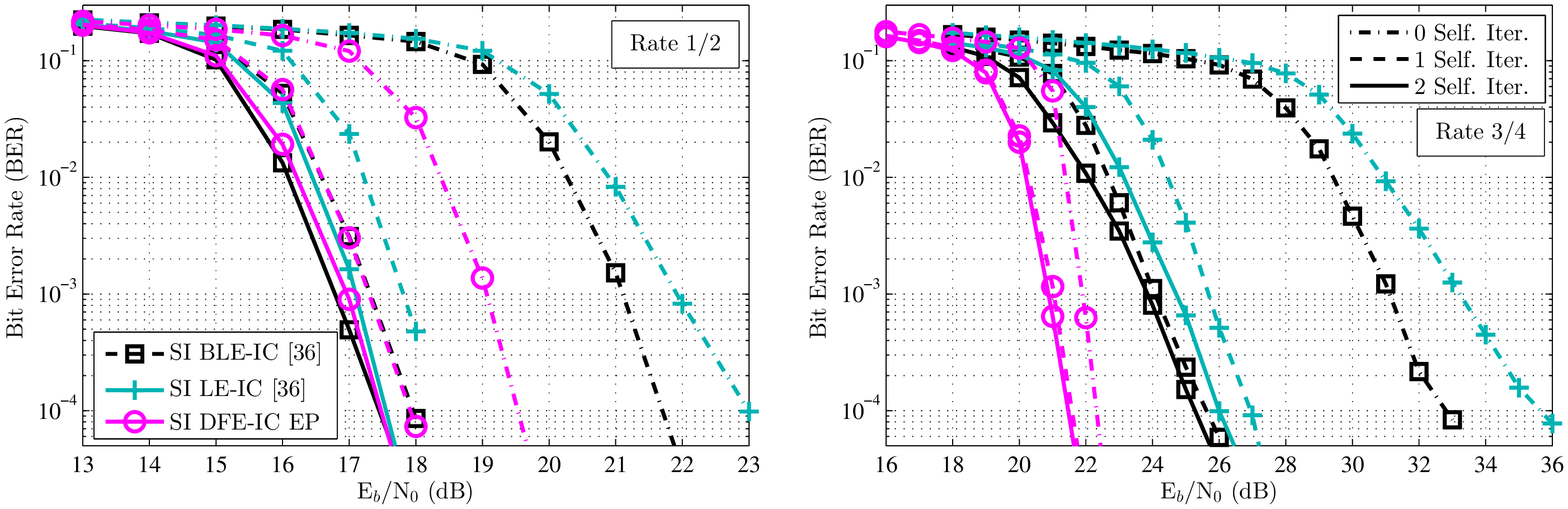}
		\caption{SI LE-IC and DFE-IC in Proakis C with LDPC coded 16-QAM, with 5 turbo iterations.}
		\label{berLDPC16QAM_proakisC}
	\end{figure*}
	\else
	\begin{figure*}[t!]
		\centering
		\includegraphics[width=6.5in]{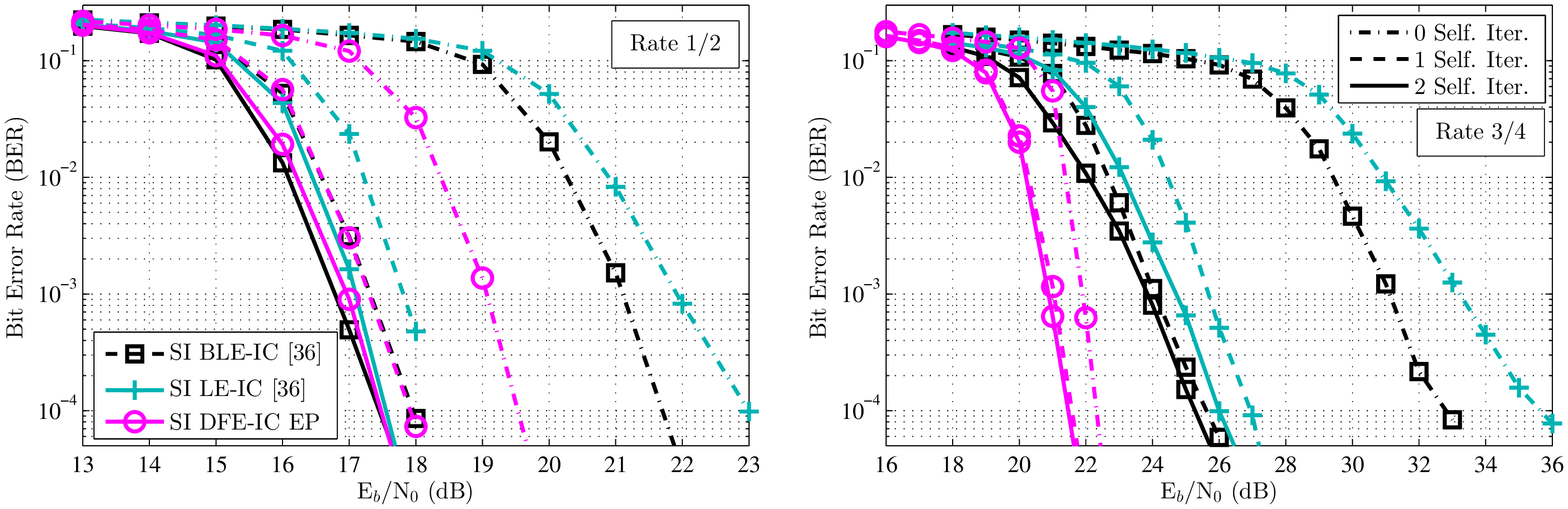}
		\caption{SI LE-IC and DFE-IC in Proakis C with LDPC coded 16-QAM, with 5 turbo iterations.}
		\label{berLDPC16QAM_proakisC}
	\end{figure*}
	\fi

	\subsection{Finite-Length Comparison}
	
	In this section, numerical finite-length results complete the previous analysis.
	In addition to receivers above, the self-iterated block linear receiver (SI BLE-IC), denoted nuBEP in \cite{santosMurilloFuentes_2017_EPnuBLE}, is considered. Without self-iterations, this receiver is equivalent to turbo block LE-IC\cite{tuchler_turbo_2011}, and it outperforms the self-iterated block receiver and Kalman smoother in \cite{santosMurilloFuentes_2017_EPBLE, santosMurillosFuentes_2017_smoorthingEP}. SI BLE-IC provides a lower bound to the BER performance of SI LE-IC.
	
	A low density parity check (LDPC) coded 16-QAM transmissions over the Proakis C channel, with rate 1/2 and 3/4 encoding of $K_b=2048$~bits (Fig. \ref{berLDPC16QAM_proakisC}).
	The proposed SI DFE-IC EP uses respectively $\beta~=~\min(0.5, 1-e^{\tau/2.5}/10)$ and $\beta~=~\min(0.1, 1-e^{\tau/1.5}/10)$ for damping, in these two cases, whereas the optimized damping reported in \cite{santosMurilloFuentes_2017_EPnuBLE} is kept for SI BLE-IC and SI LE-IC.
	The LDPC codes are obtained by path edge growth method, and a BP decoder up to a 100 iterations is used. 
	The low rate case, with (3,6) regular LDPC, shows that while all self-iterated receivers reach the same asymptotic performance as $\mathcal{S}_\tau$ increases, DFE-IC converges much faster at intermediary iterations. On the other hand, at the high rate configuration, with (3,12) regular LDPC, DFE-IC is strictly superior to LE-IC, even without self-iterations. Asymptotically even the exact SI BLE-IC is 3.8~dB behind the proposed SI DFE-IC.

	\ifdouble
	\begin{figure}[t!]
		\centering
		\includegraphics[width=3.2in]{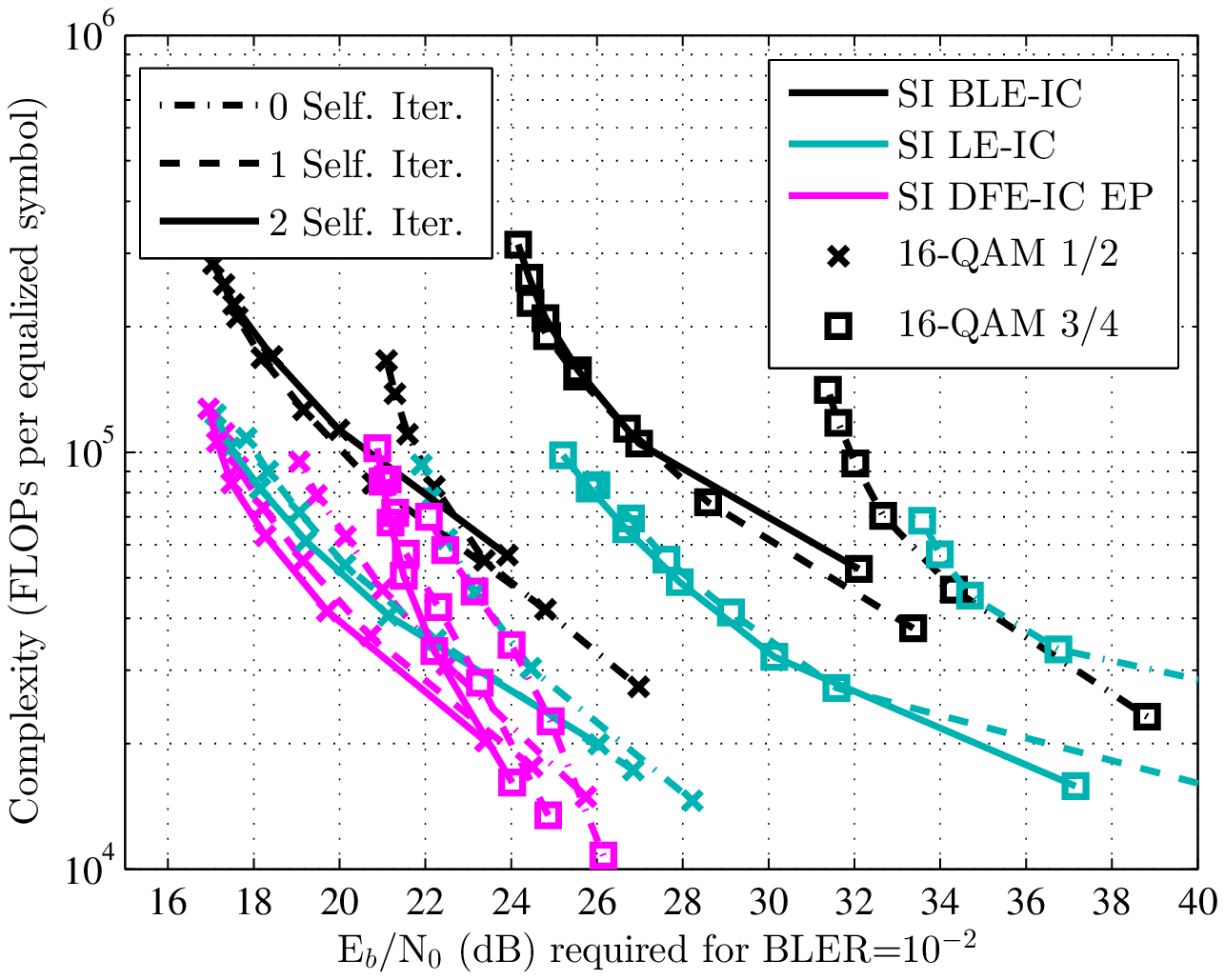}
		\caption{Performance complexity trade-off for self-iterations in LDPC coded Proakis C.}
		\label{perfLDPCComplexityTradeoff_proakisC}
	\end{figure}
	\else
	\begin{figure}[t!]
		\centering
		\includegraphics[width=3.15in]{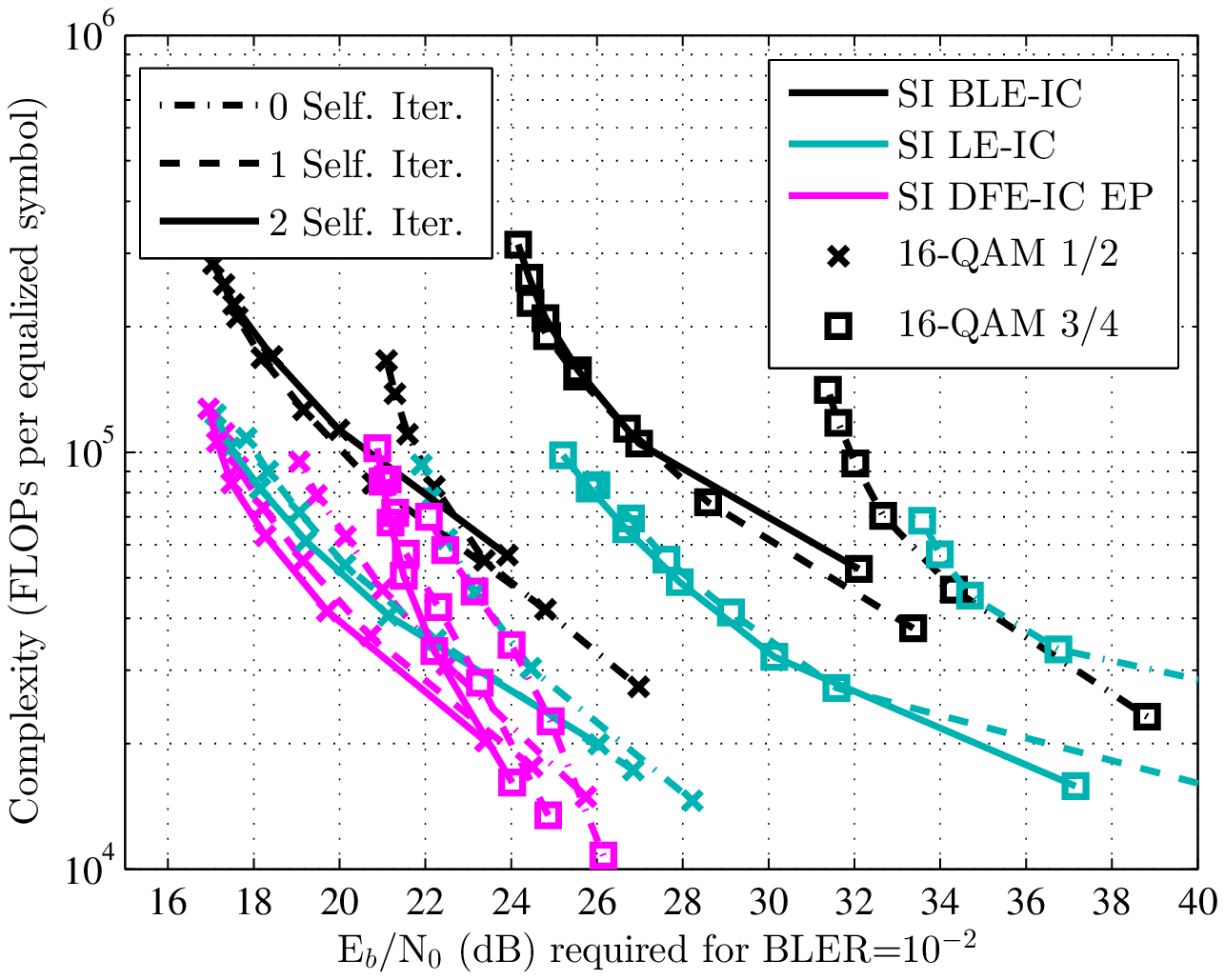}
		\caption{Performance complexity trade-off for self-iterations in LDPC coded Proakis C.}
		\label{perfLDPCComplexityTradeoff_proakisC}
	\end{figure}
	\fi

	These numerical performance results are completed with computational complexity considerations in Fig. \ref{perfLDPCComplexityTradeoff_proakisC}, where decoding threshold for $\text{BLER}~=~10^{-2}$ is evaluated for $\tau~=~0,\dots,5$, for each receiver. In the medium rate (2~bits/s/Hz: 16-QAM with rate 1/2 code) case the three considered receivers converges to the same asymptotic limit near 17~dB, but DFE-IC offers lower complexity at intermediary iterations. At 3~bits/s/Hz configuration (16-QAM with rate 3/4 code), with 5 TI and 3 SI, DFE-IC requires 3~dB less energy,  and 3 times less computational resources than BLE-IC. With $\tau=s=0$, LE-IC is unable to decode, BLE-IC decodes around 39~dB, and DFE-IC decodes with 13~dB less energy.
	
	These numerical results confirms conclusions drawn by the asymptotic analysis; the proposed SI DFE-IC is of a significant interest for high data rate applications where linear structures are less efficient. Using the efficient implementation method of section \ref{sec:matrixinvcholesk}, DFE-IC outperforms prior work in terms of both complexity and performance.
	
	\section{Conclusion}
	
	This paper investigates on the use of decision feedback with turbo equalization, for improving the limitations of linear equalizers for high data rate applications.
	
	Turbo DFE structures in the literature consist in either using hard feedback with symbol-wise adaptive filters, or soft posterior feedback with symbol-wise invariant filters. The former perform poorly at low spectral efficiency, and require complex mechanisms to improve this issue, whereas the latter are outperformed even by the conventional TV LE-IC.
	Both schemes are extended to time-variant soft feedback structures in this paper, with different filter computation hypotheses. We show that an exact approach justified with sequential Bayesian MMSE estimators (DFE-IC APP) outperforms other APP feedback alternatives. 
	
	However, due to the use of posterior estimates, this structure does not fit within the turbo principle which requires the exchange of extrinsic information. Consequently, we focus our discourse on the derivation of FIR DFE within the expectation propagation framework, which allows the computation of a novel type of extrinsic feedback from the demapper to the equalizer. 
	Building upon the emerging trend on self-iterated EP-based equalizers, the proposed DFE-IC can be self-iterated to further improve performances. 
	
	Thanks to finite-length and asymptotic analysis, DFE-IC EP, with SI or not, is shown to set new upper limits in achievable performance among FIR turbo receivers. At high data rates, even exact self-iterated block linear receivers fall over 3~dB behind the proposal.
	
	Finally, the gap of achievable rates by turbo DFE-IC to the channel capacity remains still significant at very high spectral efficiencies. Bidirectional extension of TV DFE-EP should be explored to try to close this gap.


	\appendix

	\subsection{Derivation of MMSE FIR with IC}
	\label{sec_app_mmse_fir}
	
	In this appendix, FIR equalization with interference cancellation is derived by minimizing the Bayesian MMSE criterion $J=\mathbb{E}_{\mathcal{A}_k}[|x_k-{x}_k^e{}'|^2]$, where ${x}_k^e{}'= \mathbf{f}_k'{}^T\mathbf{y}_k + g_k'$ is the equalized linear estimate, and $\mathcal{A}_k$ is a joint multivariate Gaussian prior distribution on $\mathbf{x}_k$ defined with means $\mathbf{\bar{x}}_k^{\textbf{fir}}$ and variances $\mathbf{\bar{v}}_k^{\textbf{fir}}$ (see sec. \ref{sec:ep_mmsefir}). $\mathbb{E}_{\mathcal{A}_k}[\cdot]$  and $\textbf{Cov}_{\mathcal{A}_k}[\cdot]$ respectively denote the expectation and the covariance with respect to distribution $\mathcal{A}_k$.
	Solution to this is given by $\mathbb{E}_{\mathcal{A}_k}[x_k\vert \mathbf{y}_k, \mathbf{H}_k]$, i.e. the symbol mean with respect to $p_{\mathcal{A}_k}(x_k \vert \mathbf{y}_k, \mathbf{H}_k)$.
	This distribution is the marginalization of the conjugate Gaussian posterior $p_{\mathcal{A}_k}(\mathbf{x}_k \vert \mathbf{y}_k, \mathbf{H}_k)$, i.e. of likelihood $p(\mathbf{y}_k \vert \mathbf{x}_k, \mathbf{H}_k)$ and prior $\mathcal{A}_k$. Hence, ${x}_k^e{}'$ is deduced by multiplying the MMSE estimator of $\mathbf{x}_k$ \cite{mckay_1993_fundermentals_of_stat_signal_proc} by 
	$\mathbf{e}_k$:
	\ifdouble
	\begin{eqnarray}
		\mathbf{f}_k' &=& \mathbf{e}_k^H \textbf{Cov}_{\mathcal{A}_k}[ \mathbf{y}_k, \mathbf{x}_k ] (\textbf{Var}_{\mathcal{A}_k}[\mathbf{y}_k])^{-1},  \\
		g_k' &=& \mathbf{e}_k^H \mathbb{E}_{\mathcal{A}_k}[ \mathbf{x}_k] - \mathbf{f}_k'{}^T\mathbb{E}_{\mathcal{A}_k}[ \mathbf{y}_k ],
	\end{eqnarray}
	\else
	\begin{equation}
		\mathbf{f}_k' = \mathbf{e}_k^H \textbf{Cov}_{\mathcal{A}_k}[ \mathbf{y}_k, \mathbf{x}_k ] (\textbf{Var}_{\mathcal{A}_k}[\mathbf{y}_k])^{-1},\,
		g_k' = \mathbf{e}_k^H \mathbb{E}_{\mathcal{A}_k}[ \mathbf{x}_k] - \mathbf{f}_k'{}^T\mathbb{E}_{\mathcal{A}_k}[ \mathbf{y}_k ],
	\end{equation}
	\fi
	by developing expectations above with prior statistics, it holds
	\ifdouble
	\begin{eqnarray}
		\mathbf{f}_k' &=& \bar{v}^\text{fir}_k\mathbf{h}_k^H (\mathbf{\Sigma}^\text{fir}_k)^{-1}, \\
		g_k' &=& \bar{x}^\text{fir}_k  - \mathbf{f}_k'{}^T \mathbf{\bar{x}}^\text{fir}_k,
	\end{eqnarray}
	\else
	\begin{equation}
		\mathbf{f}_k' = \bar{v}^\text{fir}_k\mathbf{h}_k^H (\mathbf{\Sigma}^\text{fir}_k)^{-1},\,
		g_k' = \bar{x}^\text{fir}_k  - \mathbf{f}_k'{}^T \mathbf{\bar{x}}^\text{fir}_k,
	\end{equation}
	\fi
	with $\mathbf{\Sigma}^\text{fir}_k = k_w\sigma_w^2\mathbf{I}_N + \mathbf{H}_k\mathbf{V}^\textbf{fir}_k\mathbf{H}_k^H$ and $\mathbf{V}^\textbf{fir}_k = \textbf{diag}(\mathbf{v}^\textbf{fir}_k)$.
	This receiver is biased, as its MMSE estimators' nature:
	\[
	\mathbb{E}_{\mathcal{A}_k}[ {x}_k^e{}' \vert x_k=x] = (1-\bar{v}_k^\text{fir} \xi_k^\text{fir}) \bar{x}^\text{fir}_k + \bar{v}_k^\text{fir} \xi_k^\text{fir} x,
	\]
	with $\xi_k^\text{fir}=\mathbf{h}_k^H\mathbf{\Sigma}^\text{fir}_k{}^{-1}\mathbf{h}_k$. 
	Removing additive and multiplicative biases with $x^e_k = (x^e_k{}' - (1-\bar{v}_k^\text{fir} \xi_k^\text{fir})\bar{x}^\text{fir}_k )/(\bar{v}_k^\text{fir} \xi_k^\text{fir})$ yields 
	the estimator given in (\ref{eq_fir_model}), which completes the proof.


	\ifCLASSOPTIONcaptionsoff
	\newpage
	\fi

	
	
	
	\bibliographystyle{IEEEtran}
	\balance
	\bibliography{IEEEabrv,bibDFE}

\end{document}